\documentclass[%
 amsmath,amssymb,
 aps,prx,twocolumn
]{revtex4-2}
\usepackage{graphicx} % Required for inserting images

\usepackage[utf8]{inputenc}
\usepackage{amstext,amssymb,amsfonts,bbm,amsthm}
\usepackage{mathtools}
\usepackage{hyperref}
\usepackage{graphicx}
\usepackage{bm}% bold math
\usepackage{xcolor}
\usepackage{booktabs} % for tables
\usepackage{siunitx} % for sci not of clustering coef
\usepackage{pgfplots}
\usepackage{tikz}
\usepackage{svg}
\usepackage{cancel}
\usepackage{braket}
\usepackage{xr} % cross-referencing between SI.tex

\newcommand{\expp}{\mathrm{exp}}
\newcommand{\squid}{\mathrm{sq}}
\newcommand{\chr}{\mathrm{chr}}
\newcommand{\start}{\mathrm{start}}
\newcommand{\nnew}{\mathrm{new}}

\newcommand{\ntot}{N_{\mathrm{tot}}}

\begin{document}

\title{Hyperdisordered cell packing on a growing surface}

\author{R. J. H. Ross}
\email{robert.ross@oist.jp}
\author{Giovanni D. Masucci}
\author{Chun Yen Lin}
\author{Teresa L. Iglesias}
\author{Sam Reiter}
\email{samuel.reiter@oist.jp}
\author{Simone Pigolotti}
\email{simone.pigolotti@oist.jp}
\affiliation{Okinawa Institute of Science and Technology, Onna, Okinawa 904-0495, Japan.
}

\date{\today}

\begin{abstract}
While the physics of disordered packing in non-growing systems is well understood, unexplored phenomena can emerge when packing takes place in growing domains. We study the arrangements of pigment cells (chromatophores) on squid skin as a biological example of a packed system on an expanding surface.  We find that relative density fluctuations in cell numbers grow with spatial scale. We term this behavior ``hyperdisordered'', in contrast with hyperuniform behavior in which relative fluctuations tend to zero at large scale. We find that hyperdisordered scaling, akin to that of a critical system, is quantitatively reproduced by a model in which hard disks are randomly inserted in a homogeneously growing surface. In addition, we find that chromatophores increase in size during animal development, but maintain a stationary size distribution. The physical mechanisms described in our work may apply to a broad class of growing dense systems.
\end{abstract}

\maketitle

\section{Introduction}

Many physical and biological systems are constituted by dense, disordered arrangements of individual units. Examples include liquids \cite{hansen2013theory}, glasses \cite{parisi2020theory}, granular systems \cite{duran2012sands}, packed macroscopic objects  \cite{renyi1963, donev2004improving, torquato2018hyperuniform}, bacterial populations  \cite{you2021confinement}, and multicellular tissues \cite{cheng2014predicting,shyer2017emergent,mongera2018fluid,gottheil2023state}. The study of these systems has led to the discovery of broad physical properties that characterize dense disordered packing.

One such property is hyperuniformity. To define it, we consider how the average number $\langle N\rangle $ of units and their variance $\sigma^2_N=\langle N^2\rangle -\langle N\rangle^2$ scale at increasing sample areas in a dense homogeneous system. For large areas, one expects a relation of the form
\begin{equation}\label{eq:numberscaling}
\sigma^2_N \sim \langle N\rangle^\alpha .
\end{equation}
If units were independently placed at random, their number in each area would follow a Poisson distribution, implying $\alpha=1$. Hyperuniform systems are characterized by having $\alpha<1$ \cite{torquato2003local}.  This means that they exhibit relative density fluctuations that are suppressed at large spatial scales \cite{zachary2011hyperuniformity, torquato2003local,dreyfus2015diagnosing}.  Originally studied in maximally random jammed systems \cite{torquato2003local,torquato2018hyperuniform}, hyperuniformity has subsequently been observed in several models in non-equilibrium statistical physics \cite{wiese2024hyperuniformity, de2024hyperuniformity, backofen2024nonequilibrium} and in active systems \cite{lei2024non}. Hyperuniformity has also been observed in biological systems, such as in the arrangement of cells in the avian retina \cite{jiao2014avian} and in leaf vein networks \cite{liu2024universal}. 

In the opposite case, $\alpha>1$, a system would display relative density fluctuations that grow with spatial scale. We term such behavior ``hyperdisordered'', although in the literature other terms such as ``anti-hyperuniform'' \citep{Torquato2021, Emmett2024} and ``super-Poissonian'' \citep{gabrielli2005statistical} are also used.  Hyperdisordered systems are characterized by long-range correlations and, in this respect, bear a similarity with critical systems in statistical physics. 
%In disordered packing, hyperdisorder appears to be less common than hyperuniformity.
In the literature to date, hyperdisordered systems have received less attention than hyperuniform ones, as the latter behavior is usually believed to be more widespread. Active physical systems in which particles form clusters characterized by a broad size distribution \cite{palacci2013living,Deseigne2010} may be considered as hyperdisordered.  Additional examples of hyperdisordered physical systems have been recently reported. These include disordered vortex matter  \cite{Puig2024}, certain regimes of active turbulence \citep{backofen2024nonequilibrium}, simulations of confined fluids \citep{leoni2024emergence}, and designed disordered systems \citep{maher2024local}.

Most theoretical ideas for dense disordered systems have been developed in the context of inanimate matter. In extending these concepts to living systems such as tissues, one should consider the effect of growth.
The impact of growth on cell arrangements remains poorly understood. Modeling shows that growth can control pattern formation and alter cell population dynamics \cite{crampin1999reaction,crampin2002pattern,krause2019influence,ross2017variable, ross2016domain}. It is thus conceivable that the presence of growth can lead to new physics in dense disordered systems. Experimentally, testing of these ideas has been limited, largely due to the difficulty in quantitatively and precisely measuring cell arrangements during growth.

In this paper,  we assess the impact of tissue growth on a dense arrangement of cells. Our experimental model system is the arrangement of pigment cells (chromatophores) on the skin of the oval squid. Our experiments reveal that chromatophore patterns are hyperdisordered. By means of a minimal model, we show that hyperdisordered behavior naturally emerges from the interplay of random cellular packing and tissue growth. Further, we find that the size distribution of chromatophores is stationary during growth. Our theory shows that this stationary distribution requires an aging mechanism, whereby individual cells possess some notion of the squid age. This prediction is in excellent quantitative agreement with our experimental measurements. Together, our results reveal fundamental physical mechanisms governing the dynamics of dense growing physical systems.

\begin{figure*}[htb]
\includegraphics[width=0.9\textwidth]{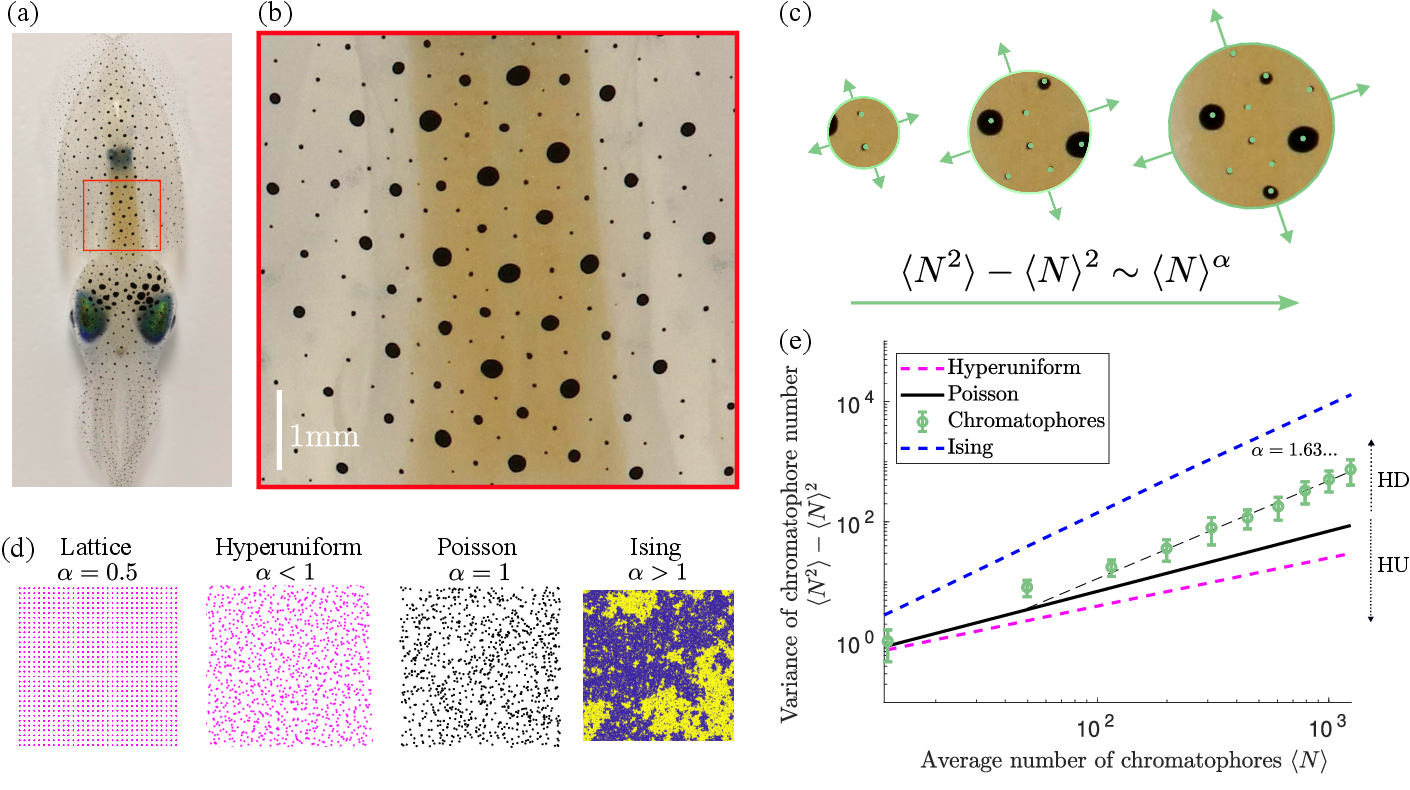}
\caption{{Chromatophore patterns are hyperdisordered.} (a) An approximately 6-weeks old squid, viewed from the top. (b) Magnification of the mantle region indicated by the red box in (a). (c)  Number of chromatophores $N$ sampled in circular areas of increasing radius 
$R$.  (d) Examples of (left to right) a uniform pattern, a hyperuniform pattern, a Poisson pattern, and a configuration of the two-dimensional Ising model at the critical temperature. (e) Chromatophore number variance, as a function of chromatophore number within an expanding area for the synthetic patterns in (d), and for squid chromatophores, which show hyperdisordered scaling (HD) (Total number of chromatophores $\ntot = 17140$ aggregated from 10 squids). A test against the null hypothesis that chromatophores form a Poissonian point pattern ($\alpha=1$) yields a p-value of $0.0026$. Here and in the following, error bars denote standard deviations, unless stated otherwise.}\label{fig1}
\end{figure*}

\section{Hyperdisordered scaling of chromatophore point patterns}

\subsection{Experimental system}

Throughout development, the skin of the oval squid  is populated by pigment-filled cells called chromatophores. Insertion of a chromatophore is thought to be possible only if the new chromatophore is at a minimum exclusion distance from preexisting ones \cite{packard1977skin, reiter2018elucidating}. Following insertion into the skin, chromatophores become attached to a collection of radially projecting muscle fibers. These muscle fibers, in turn, are innervated by neurons projecting from the brain. To perform functions related to camouflage and communication, chromatophores can be expanded beyond their resting size through neurally controlled contraction of these muscles surrounding each chromatophore \cite{messenger2001cephalopod}. However, throughout this work we used anaesthetized animals to focus exclusively on their resting size. A chromatophore's resting size (which we simply refer to as ``chromatophore size'' from now on) tends to increase as the chromatophore ages \cite{packard2011squids}.  Possibly due to their dense interactions with muscles, neurons, and other cell types, together forming the ``chromatophore organ''  \cite{Cloney_Florey1968}, chromatophores do not change their location in the skin. This means that squid chromatophores, besides being the constitutive elements of a point pattern, can also function as reference points to precisely determine skin growth \cite{packard1982chromatophores, messenger2001cephalopod}. 

To study chromatophore patterns, we took high-resolution images of the mantles of 10 different squids of the same age, see Fig.~\ref{fig1}a and ~\ref{fig1}b. The experimental protocol and imaging procedure are detailed in Appendix~\ref{app:experiments}.

\subsection{Chromatophore scaling}

Our images of squid mantles show that chromatophores are arranged in a point pattern, characterized by larger, older chromatophores being surrounded by smaller, younger ones, see Fig.~\ref{fig1}a and~\ref{fig1}b. Visually, chromatophores appear at a characteristic distance from each other, forming a lattice-like structure. The density of chromatophores in our imaging area is consistent with a uniform distribution, see Appendix~\ref{app:uniform}. 

To characterize the degree of ordering of chromatophore patterns, we measure how the chromatophore number variance $\sigma^2_N$ scales with the average number $\langle N\rangle$ of chromatophores at increasing sample areas (Fig.~\ref{fig1}c) \cite{torquato2003local}. Since chromatophore patterns (Fig.~\ref{fig1}b) emerge through packed insertion into the skin, one may expect them to exhibit hyperuniform scaling, see Fig.~\ref{fig1}d. However, since the system is growing, small gaps are constantly being created between chromatophores, resulting in cell arrangements that substantially differ from those of commonly studied jammed systems. In fact, we observe hyperdisordered behavior, i.e., 
$\alpha>1$, within the range of experimentally accessible length scales, see Fig.~\ref{fig1}e. 

\subsection{Squid model: hard disks on an expanding surface}

\begin{figure}
\includegraphics[width=8.5cm]{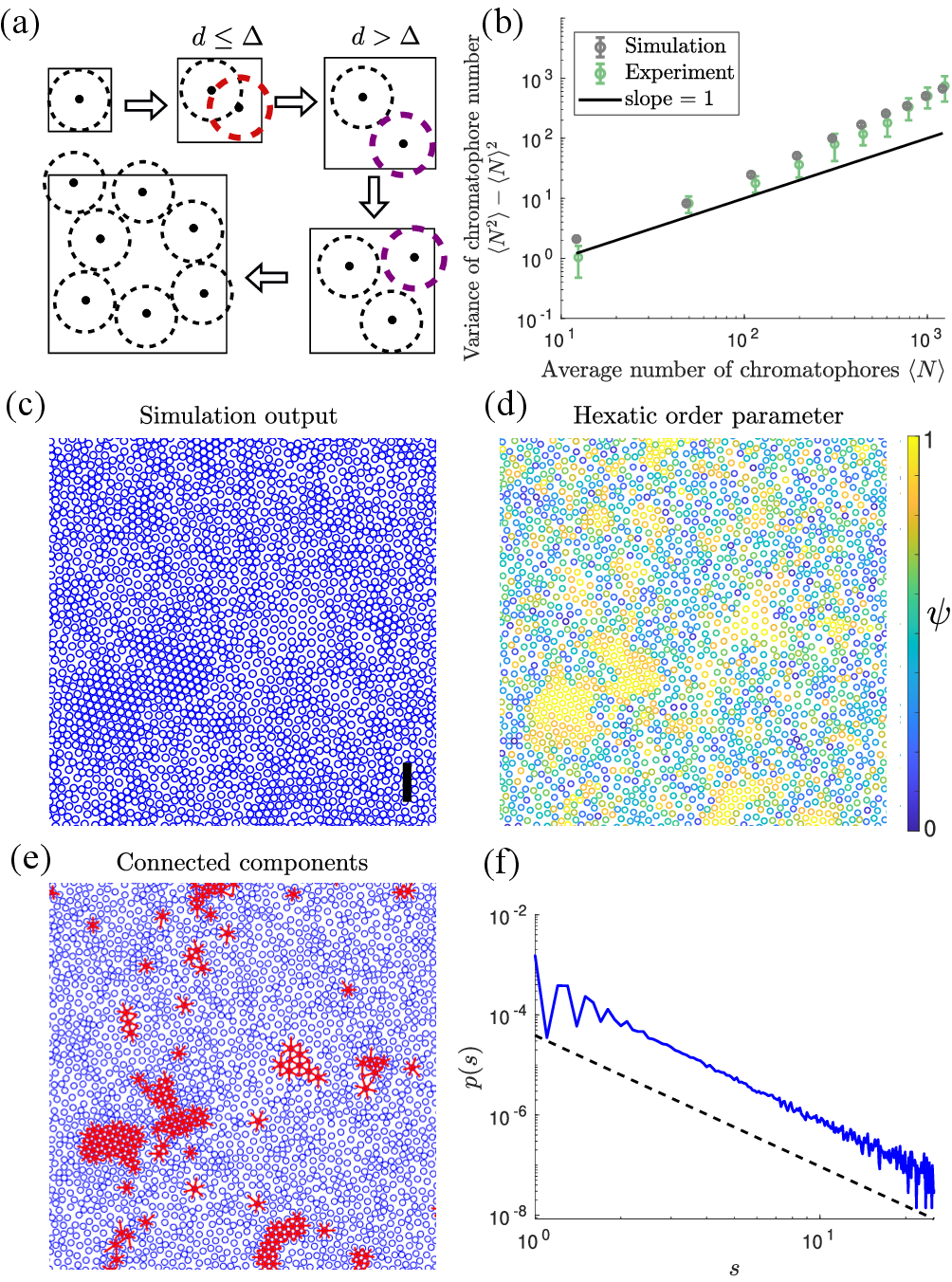}
\caption{{Growth generates hyperdisordered scaling.} (a) Schematic of the model. Black dots represent chromatophores and dashed circles their exclusion areas, which are identical for all chromatophores.  The red dashed circle represents an event in which an attempted chromatophore insertion is rejected because the distance between chromatophore centres is too small. The purple dashed circle represents a successful chromatophore insertion. (b) Number variance scaling (system size of $N=10^4$ chromatophores averaged over $500$ simulations, grey circles), compared with experimental data (green circles with error bars). (c) A configuration of the squid model (see SM Movie 1). The black scale bar indicates 1mm; the simulation area is 10mm$^{2}$.  (d) Hexatic order parameter $\psi$ for each cell in (c), see Eq.~\eqref{eq:hexatic}. (e) Red domains represent the connected components of the top 5\% cells in (d), ordered according to their value of $|\psi_j|$. (f) Size distribution of the red domains shown in (e). The black-dashed line represents a power law with exponent -2.62.}\label{fig2}
\end{figure}

To rationalize the observed hyperdisordered behavior, we introduce a model. We describe the squid mantle as a surface that grows both linearly in length and uniformly. Chromatophores are randomly placed on the mantle while the squid grows. The rate of attempted chromatophore insertion is proportional to the surface area, and the location of attempted insertions is uniformly chosen on the surface. An attempted chromatophore insertion is successful only if the distance between the candidate new chromatophore and all the existing ones is larger than an exclusion distance $\Delta$, see Fig.~\ref{fig2}a. The model is therefore equivalent to a system of randomly placed hard disks on a homogeneously growing surface. The linear growth rate of the surface and the exclusion distance are determined by matching the surface growth rate and spatial density of chromatophores from experiments. Additional details on the model simulations and parameter choices are presented in Appendix~\ref{app:squidmodel}. Despite its simplicity, the model generates patterns presenting hyperdisordered scaling of the number variance, consistent with that observed in real squids (Fig.~\ref{fig2}b).

We next seek to understand the cause of this scaling behavior. A large-scale simulation of this model shows nearly regularly packed domains, with different densities and without a clear characteristic size, surrounded by more disordered regions, see Fig.~\ref{fig2}c. During growth, the density of these domains oscillates in a sawtooth manner. The reason for this is that the domains gradually become less packed as the surface grows, until there is enough space for inserting new chromatophores in the gaps between existing ones, see SM Movie 1.  To quantify the statistics of these domains, we introduce the hexatic order parameter\begin{equation}\label{eq:hexatic}
\psi_j=\frac{1}{\nu_j}\sum_{l=1}^{\nu_j} e^{i6\theta_{jl}}\, ,
\end{equation}
where $j$ is an index representing a given cell, $\nu_{j}$ is the number of nearest neighbours of cell $j$ computed via Delauney triangulation, and $\theta_{jl}$ is the angle between the vector pointing from cell $j$ to $l$ and the $x$ axis. A large value of $|\psi_i|$ can be interpreted as a local ordering of the neighbors of cell $j$, see Fig.~\ref{fig2}d. We therefore identify domains as connected sets of neighboring cells characterized by high values of $|\psi_j|$, see Fig.~\ref{fig2}e. We find that the size distribution of these domains is well described by a power law (Fig.~\ref{fig2}f), consistent with the idea that the dynamical heterogeneity generated by the interplay of growth and area exclusion is scale free.

\begin{figure}[htb]
\includegraphics[width=8.5cm]{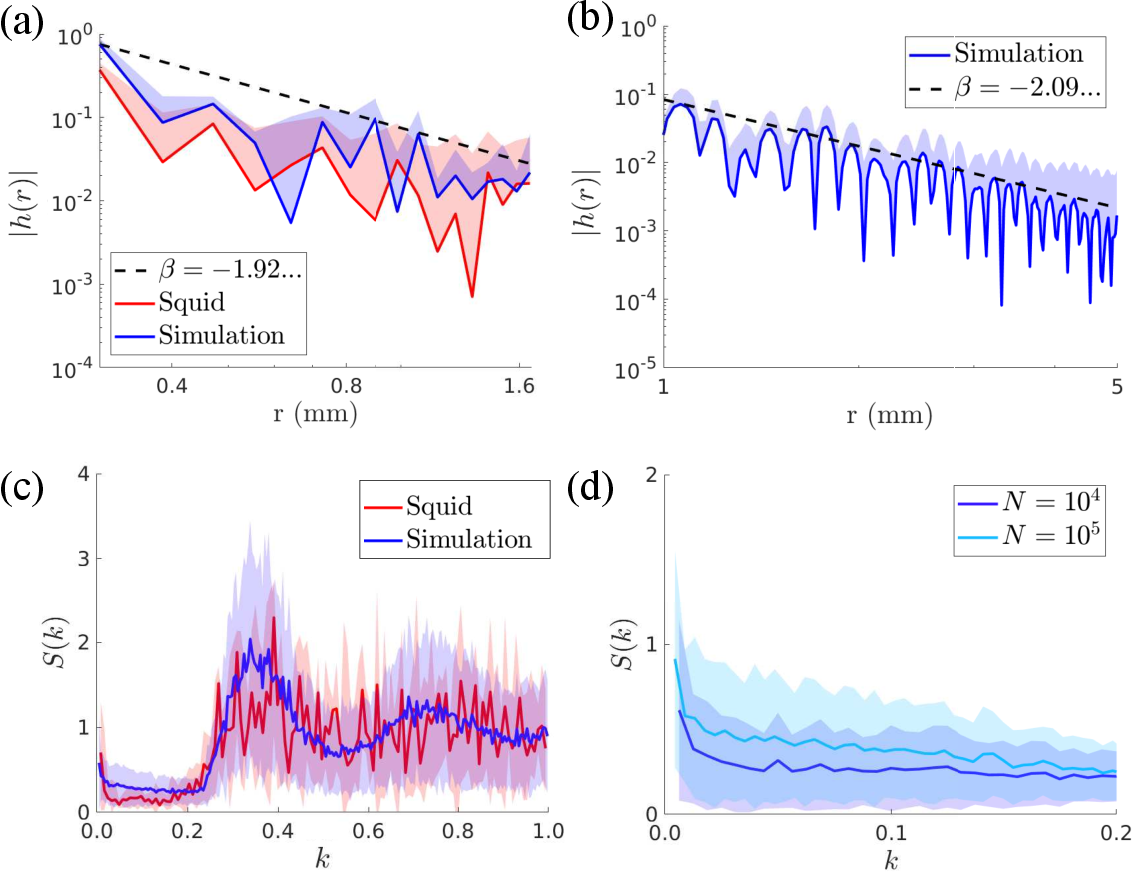}
\caption{Correlation functions and structure factors are consistent with hyperdisordered behavior. (a) Comparison of correlation functions from experiment and simulation of the squid model. Here, simulations are performed up to a comparable system size as the experimental system ($N=2\cdot 10^3$).  The correlation function is defined as $h(r) = \rho_{2}(r)/\rho^{2} - 1$, where $\rho_{2}(r)$ is the density of pairs at distance $r$ and $\rho$ is the one-body density. The black dashed line is a fit of the envelope of the simulation data. Here and in the following, shaded regions denote standard deviations. (b) Correlation function in the squid model for a system size much larger than in the experimental system ($N=2\cdot 10^4$). (c) Comparison of structure factor for squid chromatophores and from simulation data. In this case, the final system size is $N=10^4$ chromatophores averaged over $500$ simulations. (d) Structure factors from the squid model for different system sizes.} \label{fig3}
\end{figure}

Hyperdisordered scaling of a point pattern is necessarily associated with a pair correlation function whose integral in space diverges, see Appendix~\ref{app:triad} and Ref.~\cite{torquato2018hyperuniform}. The envelope of the correlation function, both in the squid and in the model, appears to decay as 
$r^{-2}$, see Fig.~\ref{fig3}a. This decay implies a (logarithmic) divergence of the integral of the correlation function in two dimensions, as expected.
The $r^{-2}$ decay is more evident in a large-scale simulation of the model, see Fig.~\ref{fig3}b. 

Another equivalent property of a hyperdisordered system is that the structure factor 
\begin{equation}\label{eq:structurefact}
S(k)= \frac{1}{N}\sum_{j,l=1}^N e^{i\vec{k}\cdot(\vec{x}_j-\vec{x}_l)}
\end{equation}
must diverge in the limit $k\rightarrow 0$, see Appendix~\ref{app:triad}. This behavior contrasts with that of hyperuniform systems, in which $S(k)\rightarrow 0$ in the same limit (see Appendix~\ref{app:triad}). This property is difficult to test in practice, as taking this limit requires a very large system. The structure factor measured from our experimental data presents an increasing behavior as $k\rightarrow 0$, see Fig.~\ref{fig3}c. A similar behavior is observed in the model. Running simulations of the model for larger system size reveals a more pronounced peak of the structure factor near the origin, see Fig.~\ref{fig3}d, as expected for a hyperdisordered system.

\begin{figure}[htb]
\includegraphics[width=7.5cm]{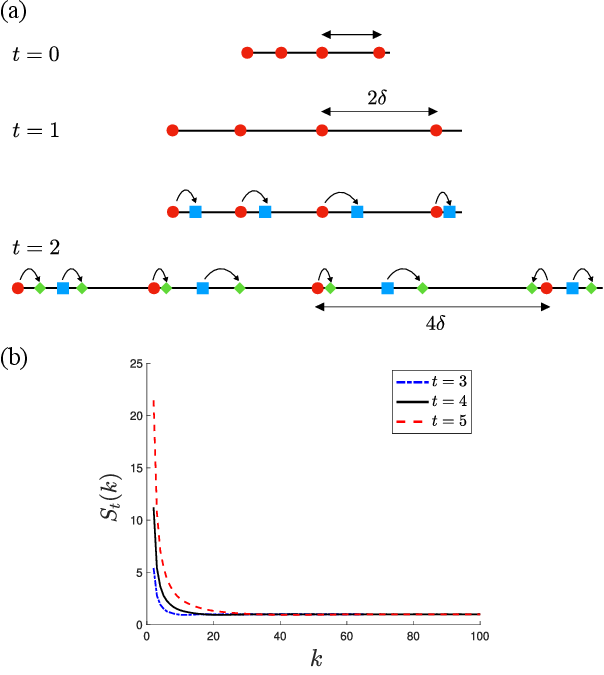}
\caption{An iterative model presents hyperdisordered behavior. (a) Schematic of the model. (b) Structure factor as expressed by Eq.~\eqref{eq:recurrence_formula} for $L=4$, $\sigma^{2} = 1$, and $t = 3, 4, 5$ as shown in the legend. \label{fig4}}
\end{figure}

\subsection{One-dimensional iterative model}

Why does growth lead to hyperdisordered behavior in a dense system? To gain understanding, we propose a one-dimensional iterative model that recapitulates some of the salient features of our system. In this model, a 1D domain is initialized with $n$ cells placed uniformly at random. In analogy with the squid model, we call $\Delta$ the typical distance between cells. We set $\Delta = 1$, so that an initial domain of integer length $L$, contains $n= L$ initial cells. At each step of the model (generation), the domain size is doubled and new cells are placed at $x_{i} + \delta_x(\Delta,\sigma^{2})$, where $x_{i}$ are the positions of the cells already present and $\delta_x$ is a Gaussian random variable with mean $\Delta$ and variance $\sigma^{2}$ (Fig.~\ref{fig4}a). This process is repeated for $t$ steps, so that the final domain size is $n\times 2^t$.

To analyze the model, we study the structure factor $S_t(k)$ of the model at generation $t$. We recall that a point pattern is hyperdisordered if the structure factor diverges as $k\rightarrow 0$. We find that the structure factor is expressed by
\begin{align}\label{eq:recurrence_formula}
S_{t}(k) &= 2^{t}S_{0}(2^{t}k)\prod_{i=1}^{t}F(2^{t-i}k) \nonumber \\ &+ \sum_{i=1}^{t}2^{t-i}\phi(2^{t-i}k)\prod_{j=1}^{t-i}F(2^{t-i-j}k),
\end{align}
where
\begin{equation}\label{eq:F}
F(k) = \frac{1}{4}\left(1 + 2\cos(\Delta k)e^{-\frac{k^{2}\sigma^{2}}{2}} + e^{-k^{2}\sigma^{2}}\right),
\end{equation}
and
\begin{equation}\label{eq:phi}
\phi(k) = \frac{1}{2}\left(1 - e^{-k^{2}\sigma^{2}}\right),
\end{equation}
see Appendix~\ref{app:iterative} for a derivation. In the limit of large $t$ and  $k\rightarrow 0$, the structure factor given in Eq.~\eqref{eq:recurrence_formula} diverges (see  Fig.~\ref{fig4}b), implying that the behavior of the iterative model is hyperdisordered, regardless of parameter choice. This can be shown by setting $k = \left(2 \pi /(n 2^{t} \right))$ in Eq. \eqref{eq:recurrence_formula}. We then use that, for some $i < t$, we have  
\begin{align}\label{eq:Flimit}
F\left(\frac{2^{t-i} 2 \pi}{n 2^{t}}\right) &= 1 + 2\cos\left(\frac{\Delta 2 \pi}{n 2^{t}}\right)e^{-\frac{1}{2}\left(\frac{2^{t-i} 2\pi \sigma}{n 2^{t}}\right)^{2}} \nonumber\\
&+ e^{-\left(\frac{2^{t-i} 2\pi \sigma}{n 2^{t}}\right)^{2}} > 2.
\end{align}
This result implies
\begin{equation}
\lim_{t\rightarrow\infty}S_{t}(2\pi/(n 2^{t})) \rightarrow \infty.
\end{equation}

The results of the iterative model suggest that, in essence, hyperdisordered behavior is caused by growth exporting density fluctuations to increasingly large scales.

\section{Dynamics of chromatophore size presents aging}

\begin{figure*}[htb]
\includegraphics[width=0.9\textwidth]{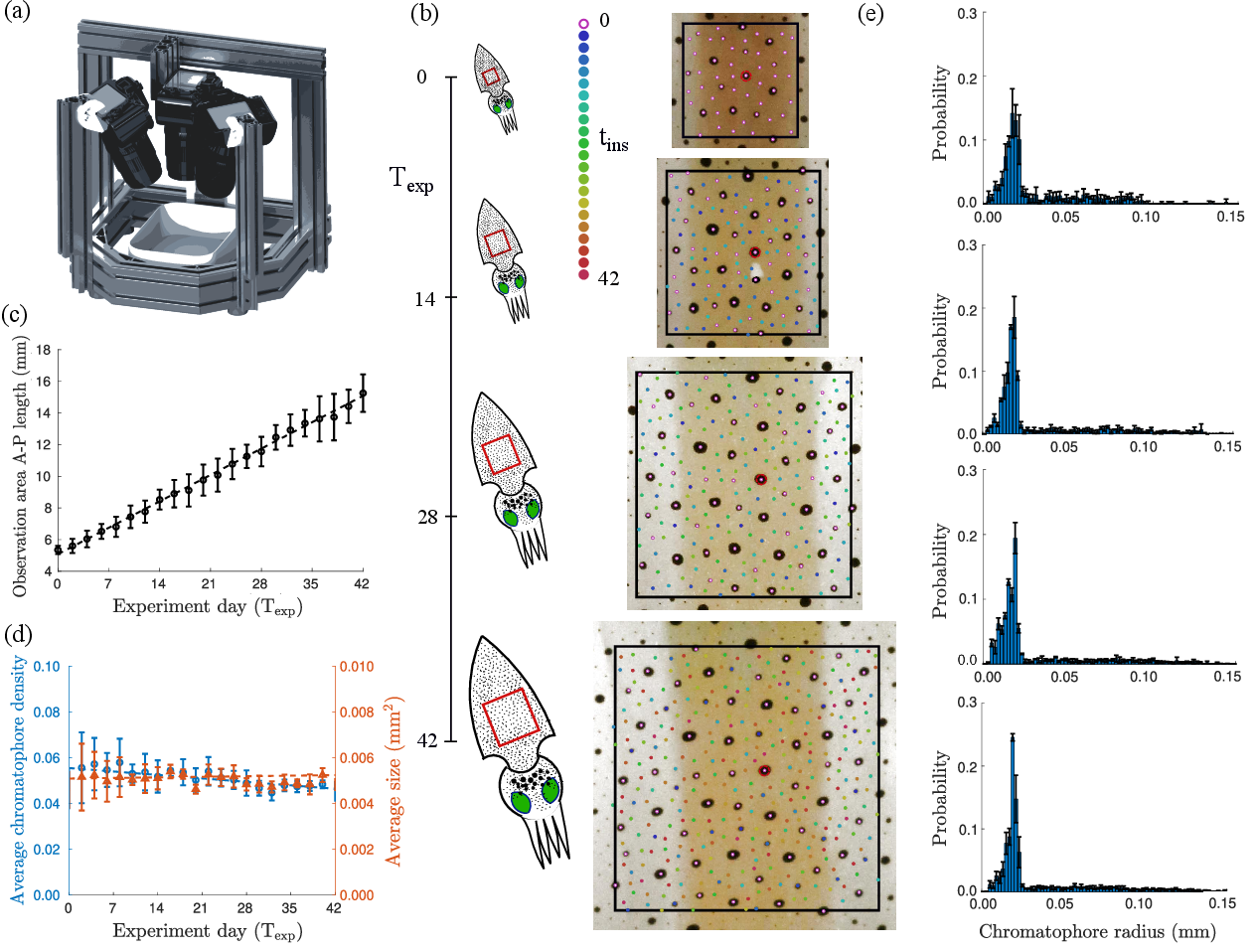}
\caption{Chromatophore size distribution is stationary during growth. (a) Experimental setup for 3D imaging of squid skin. (b) Left: schematic of chromatophore tracking experiment. Right: skin patch with tracked chromatophores at experiment day $(\mathrm{T_{\expp}}$) 0, 14, 28, and 42. Color denotes chromatophore age. (c) Average observation area anterior-posterior length for three squids over 6 weeks.  Dashed black line is a linear fit to experimental data (5.08 + 0.24 $T_{\expp}$). (d) Average density and size of chromatophores in the observation area for three squids over 6 weeks. Dashed blue and orange lines are linear fit to correspondent experimental data (0.05 - $2\cdot 10^{-4}T_{\expp}$, and $5 \cdot 10^{-3} + 4 \cdot 10^{-6}T_{\expp}$, respectively). (e) Distribution of chromatophore radii for day $(T_{\expp}$) 0 ($\ntot=706$), 14 ($\ntot=1591$), 28 ($\ntot=2715$) and 42 ($\ntot=5029$) from 3 squids. \label{fig5}}
\end{figure*}

\subsection{Stationary distribution of chromatophore size}

We now study the dynamical behavior of chromatophore patterns, in particular how chromatophore sizes evolve as the system grows. To track populations of chromatophores on growing patches of squid skin through time, we employed three-dimensional imaging to construct the skin manifold (Fig.~\ref{fig5}a), and used computer vision techniques to segment chromatophores and locate their center of mass, see Fig.~\ref{fig5}b and Appendix~\ref{app:experiments}.  Tracking an initial set of chromatophores over 6 weeks reveals uniform linear growth in the inter-chromatophore distance over time, and consequently also in the system size (Fig.~\ref{fig5}c). In parallel, individual chromatophores grow in size as they age (Fig.~\ref{fig5}b). Increases in chromatophore radii and separation are offset by the insertion of small new chromatophores into the skin, resulting in an approximately constant chromatophore density and stationary size distribution over time, see Fig.~\ref{fig5}d and ~\ref{fig5}e. The stationarity of the chromatophore size distribution is confirmed by a Kolmogorov-Smirnoff test, see Appendix~\ref{app:KS}.

\subsection{Aging in chromatophore growth dynamics}

A stationary chromatophore size distribution over development implies that chromatophores must grow in size at a rate that depends on squid age. We refer to this phenomenon as ``aging'', in analogy with materials such as plastics whose physical properties depend on time.

\begin{figure*}[htb]
\includegraphics[width=0.9\textwidth]{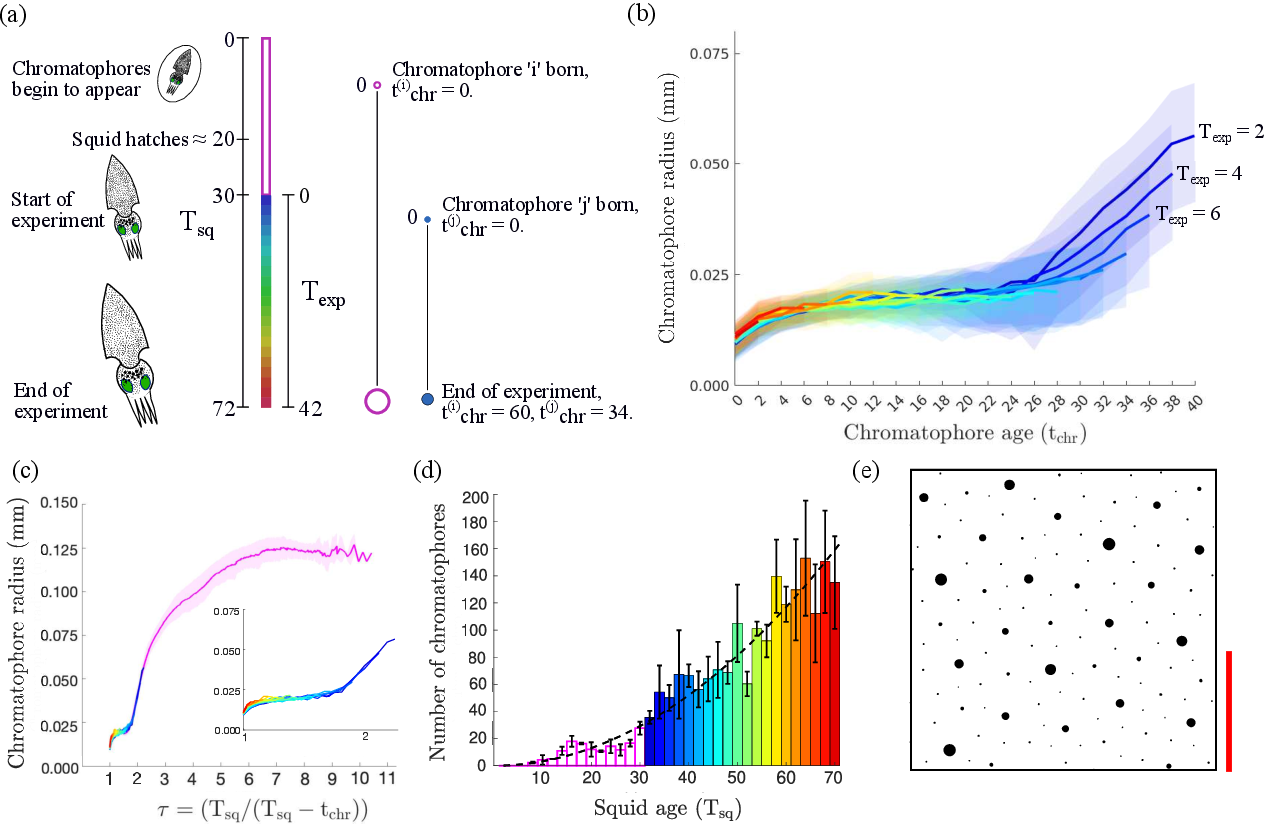}
\caption{{Chromatophore growth rate decays as the squid ages.} (a) Timeline of squid development, where $T_{\squid}$ is the time measured from the fertilization of the squid egg, $T_{\expp}$ the time measured from the start of the experiment, and $t_{\chr}^{(j)}$ the age of chromatophore $j$. (b) Average chromatophore radius as a function of chromatophore age. Color denotes the experimental time, $T_{\expp}$, of chromatophore insertion ($\ntot= 5733$ from 3 squids followed from approximately 2-weeks-old to 8-weeks-old). (c)  Average evolution of chromatophore radii as a function of chromatophore age with pre-experiment chromatophores included ($\ntot = 6183$ from 3 squids). (inset) Chromatophore radii from (b) plotted as function of $\tau$. Inset axes are same as (c). (d) Insertion times of all chromatophores. The black dashed line represent a quadratic fit. (e) Chromatophore pattern from simulation. The simulation area displayed is 2.5mm$^{2}$. The red scale bar indicates 1mm.\label{fig6}}
\end{figure*}

To prove that the stationarity of chromatophore size distribution requires aging, we call $T_{\squid}$ the age of the squid, measured from the fertilization time of the egg (Fig.~\ref{fig6}a). We also define the time $T_{\expp}$ elapsed from the first experimental observation, and the age $t_{\chr}^{(i)}$ of a given chromatophore, labeled by index $i$ (Fig.~\ref{fig6}a). Given that the mantle grows linearly  (Fig.~\ref{fig5}c) and that the density of chromatophores is approximately constant during growth, the average total number of chromatophores $\mathcal{N}(T_\squid)=\langle N\rangle$ at squid age $T_\squid$ is proportional to the mantle area, which in turn grows as the square of the squid age, $\mathcal{N}(T_\squid)\propto T_\squid^2$. The distribution $g(t_{\chr})$ of chromatophore ages $t_{\chr}$ for $T_\squid\gg 1$ is proportional to $\dot{\mathcal{N}}(T_\squid - t_\chr)$, i.e., how many chromatophores were inserted at time $(T_\squid - t_\chr)$. After normalization, we conclude that $g(t_{\chr})=2(T_\squid-t_\chr)/T_\squid^2$, for large $T_\squid$.

We call $p(R|t_\chr,T_\squid)$ the probability that a chromatophore has radius $R$, given its age $t_\chr$ and the squid age $T_\squid$. If the radius is uniquely determined by these two quantities, this probability is given by a delta function. We similarly define the probability $p(R)$ that a randomly chosen chromatophore has radius $R$. We obtain
\begin{equation}\label{eq:convol}
p(R)=\int_0^\infty d t_\chr \, p(R|t_\chr,T_\squid) \frac{2(T_\squid-t_\chr)}{T_\squid^2} .
\end{equation}
If $p(R|t_\chr,T_\squid)$ did not depend on $T_\squid$, then Eq.~\eqref{eq:convol} would imply that $p(R)=2[T_{\squid}f_1(R)-f_2(R)]/T_{\squid}^2$, with $f_1(R)=\int_0^\infty dt_{\chr} p(R|t_{\chr})$ and  $f_2(R)=\int_0^\infty dt_{\chr} p(R|t_{\chr}) t_{\chr}$. Such $p(R)$ would explicitly depend on $T_\squid$ and would therefore be non-stationary.

This argument implies that aging is necessary to achieve a stationary size distribution, as anticipated. In particular, if $R$ is a deterministic function of $\tau=T_\squid/(T_\squid-t_\chr)$, then the distribution $p(R)$ would always be stationary. This result can be verified by directly substituting $p(R|t_\chr)=\delta(R-f(\tau))$ into Eq.~\eqref{eq:convol}, where $f(\tau)$ is an arbitrary increasing function. In summary, our theory predicts that the rate of chromatophore growth must slow as the squid ages. 

In agreement with our theoretical prediction, we find that, at equal chromatophore ages, chromatophores that are born earlier tend to be larger (Fig.~\ref{fig6}b). In contrast, when plotting chromatophore radii as a function of $\tau$, we obtain a precise collapse to a master curve (Fig.~\ref{fig6}c). In this collapse, we treat $T_\squid$ as a free parameter, permitting us to estimate the value of $T_{\squid}$ corresponding to the start of the experiment, i.e., $T_{\expp} = 0$. This value is $T_{\squid} = 30.7$, which is consistent with the sum of the pre-hatching time (approximately 20 days) and the time between hatching and the beginning of our experiment (approximately 14 days). Further, by fitting individual chromatophore growth curves to the master curve, we estimate the insertion date of chromatophores that appeared before the start of our observations, see Fig.~\ref{fig6}c and Appendix~\ref{app:insertiontimes}. The complete distribution of chromatophore ages is well fitted by a quadratic law (Fig.~\ref{fig6}d, $r^{2} = 0.93$). The predicted time for the appearance of the first chromatophore is $T_{\squid} = 6.4$, which is consistent with the time of appearance of the first chromatophores on the unhatched squid after fertilization \cite{samuel2002intercapsular}. Thus, individual chromatophores reduce their growth rate as the squid ages in such a way that the distribution of chromatophore sizes is stationary during growth.

In our model, we have assumed for simplicity that the exclusion zone associated with a chromatophore is independent of its size. This assumption permitted us to study the point pattern and the chromatophore size distribution as two independent physical problems, and leads to predictions that are in quantitative agreement with our experimental observations. Despite this independence, the model is able to produce patterns in which large chromatophores tend to be surrounded by smaller ones (Fig.~\ref{fig6}e), also in agreement with experimental observations, see Fig.~\ref{fig1}a.

\section{Discussion}

In this work, we have demonstrated how growth generates a hyperdisordered scaling of squid chromatophore patterns. By combining experimental measurements and theory, we revealed that this unconventional state of matter spontaneously arises through the interaction of random packing and growth. In particular, growth is responsible for exporting short-range disorder generated by the packing process to ever larger spatial scales. This scaling behavior is quantitatively captured by a simple system of static disks randomly placed on a growing surface, in which growth generates scale-free dynamical heterogeneity. Given the simplicity of this mechanism, we expect that hyperdisordered behavior might be observed in other growing physical or biological systems. In contrast, it has been observed that photoreceptors form a pattern on the chicken retina which possesses hyperuniform, rather than hyperdisordered, scaling properties \cite{jiao2014avian}. It has been hypothesized that this behavior may provide optimal retinal coverage properties for vision. This suggests that, depending on the interaction of cell dynamics and tissue growth, biological point patterns can be funneled into different kinds of scaling behavior. 

By tracking the position of chromatophores as the squid grows, we have shown that the macroscopic growth of the squid mantle can be idealized as a uniformly expanding surface.  However, at the cellular level skin growth and chromatophore insertion in squid is likely mediated by stem cell division and differentiation, as in other animal epithelial tissues \citep{chan2022skin, dekoninck2020defining, itzkovitz2012optimality}. Such cellular integration involves, to various degrees, cell migration \citep{deryckere2021identification, garcia2020impact}.  At the tissue level, these
dynamics can be characterized by jamming, reminiscent of disordered glassy materials  \citep{mongera2018fluid, lenne2022sculpting}. Future mechanistic studies could elucidate how these effects shape the dynamics of the squid mantle.

Some tissues other than the squid mantle are also characterized by exclusion zones.  Our observation of exclusion zones shaping the patterning of chromatophores, where cells maintain a minimum distance from each other, resembles the situation in other cephalopod species \cite{reiter2018elucidating}, as well as in hair follicles \cite{cheng2014predicting} and feather buds \cite{shyer2017emergent}. It also is reminiscent of plant growth, which is constrained by the mechanical properties of cell walls \citep{coen2023mechanics}. Similarly to the squid, the displacement field associated with growth in planar plant leaves is angle-preserving, implying that growth is locally isotropic \citep{alim2016leaf}.  An exciting future possibility is to experimentally manipulate the growth of the squid to assess its effect on the chromatophore pattern and associated density fluctuations. 
 
It is natural to speculate whether hyperdisordered scaling serves a biological purpose. One hypothesis is that this type of patterning facilitates the camouflage or communication functions of the squid skin display system. Indeed, it has been shown that systems displaying fluctuations on a broad range of scales are particularly efficient at processing complex environmental signals \cite{hidalgo2014information}. Further, recent work on plant morphogenesis has suggested that the exporting of fluctuations to large scales via growth may increase developmental robustness \cite{fruleux2024growth}.  Alternatively, hyperdisordered scaling might be an evolutionary spandrel \cite{gould1979spandrels}, i.e. a side effect of the interaction of growth and packing.   

Additionally, we have found that the distribution of chromatophore sizes is maintained over time by slowing the rate of chromatophore growth as the squid ages. This result implies that chromatophores must possess some notion of squid age during growth. How this knowledge is acquired remains unclear. One possibility is that chromatophore growth rates are controlled by a morphogen that scales with system size \cite{wartlick2011dynamics, averbukh2014scaling, aguilar2018critical}. Another is that age-dependent mechanical forces modulate chromatophore growth, as they regulate growth in plants \citep{boudaoud2010introduction, uyttewaal2012mechanical, hamant2008developmental} and animals \citep{shraiman2005mechanical, lecuit2007orchestrating}. A third possibility is hormonal regulation \citep{shi2000amphibian, salis2021thyroid}.  Our conclusion that chromatophore dynamics must present aging is based on a simple dimensional argument. This argument may therefore be applied to other dynamical processes on linearly growing surfaces that generate stationary patterns.

In summary, our work illustrates that the interplay of dynamics and tissue growth can unlock patterns that would be impossible in non-growing tissues. Given the ubiquity of this interplay, we expect this idea to extend far beyond our particular model system.

\begin{acknowledgements}
We are grateful to the OIST Cephalopod Research Support Team and Zdenek Lajbner for animal care, for the help and support provided by the Scientific Computing and Data Analysis and Engineering sections, Research Support Division at OIST. We thank all Pigolotti and Reiter lab members for assistance and discussion. We thank Mahesh Bandi, Joshua Shaevitz, and Salvatore Torquato for feedback on a preliminary version of this manuscript.
\end{acknowledgements}

\appendix

\section{Experimental system, imaging, and chromatophore tracking}\label{app:experiments}

We here briefly describe our experimental model system and the imaging process. Oval squids (\textit{Sepioteuthis   lessoniana}) were bred in the OIST Marine Science Station. Juvenile squids used for this study were group housed in 50x40x40cm tanks connected to the ocean through an open seawater circulation system.  The squids grow rapidly, hatching with a mantle length of approximately 16mm and reaching 90mm within 3 months. All research and animal care procedures were carried out in accordance with institutional guidelines, approved by the OIST Animal Care and Use Committee under approval number 2019-244-6.

For the analysis presented in Fig. \ref{fig1}, 10 animals (8 weeks post-hatching) were anesthetized (1 percent ethanol \cite{butler2018vivo}), and individually transferred to a filming chamber for approximately 1 minute (Fig.~\ref{fig2}a). They were photographed using a high resolution camera ($8688 \times 5792$ pixels, approximate pixel size 7$\mu$m $\times$ 7$\mu$m). After imaging, animals were transferred to filtered seawater, and returned to their home tank after waking from anesthesia.

For chromatophore tracking, 12 animals (2-3 weeks post-hatching) were photographed in three dimensions using a 3-camera rig (Fig.~\ref{fig2}a). Image acquisition was synchronized via an Arduino. Cameras were calibrated using a checkerboard (square size = 5mm $\times$ 5mm), and the reprojection error was typically less than a pixel. Images were taken every 2 days for 42 days. Squid identity was maintained across days using the spatial arrangement of chromatophores.  

Within an image, chromatophore locations were determined through segmenting chromatophores using a U-net \cite{woo2023dynamics,ronneberger2015u}, followed by image binarization and centroid detection. These locations were manually curated via visual inspection. 

For multi-camera imaging, points corresponding to chromatophore centroids were matched through nonlinear warping of pairs of images using a custom GUI \cite{woo2023dynamics}, followed by nearest-neighbour association of points across the two images. With chromatophore centroids identified across simultaneously acquired images, the three-dimensional location of chromatophores within a day was determined using the camera calibration.  Chromatophores were tracked across days in a similar manner, using nonlinear warping and nearest-neighbor matching between images separated by 2 and 4 days. A chromatophore was labelled as new if it did not have a corresponding chromatophore from the previous 2 and 4 day images.

\section{Uniformity of the spatial distribution of chromatophores}\label{app:uniform}

We divided equally sized images of chromatophores into 16 equally sized sections and we compared the number of chromatophores in each box to both the expected number of chromatophores within an image, and across all images (total number of images $= 10$), using a $\chi^{2}$ test. The  hypothesis that chromatophores were present in equal proportions across sections, was not rejected in all cases at significance level $\beta = 0.05$.  In particular, we obtained values of the $\chi^{2}$-statistics (5.426, 9.425, 3.304, 9.514, 7.594, 2.822, 5.024, 15.790, 16.593, 5.712), corresponding to p-values  (0.988, 0.854, 0.999, 0.849, 0.934, 0.999, 0.992, 0.396, 0.344, 0.984). We conclude that the spatial distribution of chromatophores is uniform, within our experimental uncertainty.

\section{Squid model}\label{app:squidmodel}

Here, we provide details on the squid model. Chromatophore insertion is attempted by drawing new coordinate pairs ($x_{\nnew}$, $y_{\nnew}$) uniformly and at random, from within the current domain. Chromatophore insertion is successful if
\begin{equation}
(x_{\nnew}-x_i)^2 +(y_{\nnew}-y_i)^2 > \Delta^{2}\qquad \forall i ,
\end{equation}
where $(x_i,y_i)$ are the coordinates of an existing chromatophore $i$ and the parameter $\Delta$ is the exclusion distance, i.e., the minimum distance between chromatophore centers. Otherwise, the attempt is discarded. 

During the simulation, chromatophore insertions are repeatedly attempted, until a maximum number of sequential failures is reached. Once this occurs, the domain length in both the $x$ and $y$ direction is increased by $P_d\, dt$, where $P_d$ is the linear growth rate and $dt$ is the simulation time step. In this growth step, the coordinates of all chromatophores are rescaled proportionally to the new domain size. Chromatophore insertion is again attempted until the maximum number of sequential failures is reached. The maximum number of sequential failures $f$ is increased quadratically in time throughout the simulation, $f = 5(6.4 + t^{2})$, to maintain a fixed number of failed attempt per unit area.

As for the parameters, the exclusion distance is set to $\Delta = 0.25$ mm, chosen in order to match the experimental chromatophore density (see Fig.~\ref{fig5}d).  The initial domain area is 1.5$\Delta \times 1.5\Delta$. The domain linear growth rate is $P_{d} = 0.24$ mm per day (see Fig. \ref{fig5}c).  The initial simulation time is $t=T_{\start} = 6.4$ days, in accordance with the time of first appearance of chromatophores inferred from experimental data (see Fig. \ref{fig6}). We set a timestep equal to $dt=10^{-3} \mbox{days}$. All simulations were written in C++ and run on the supercomputer Deigo at OIST.

\section{Relation between the scaling of the number variance, the structure factor, and the total correlation function}\label{app:triad}

In this Appendix, we outline the relationship between the scaling of the number variance, the total correlation function, and the structure factor, and how this can be used to distinguish between Poisson, hyperuniform, and hyperdisordered point patterns. Our derivation closely follows references \citep{torquato2003local, torquato2018hyperuniform}.  
% The pair correlation function is defined as as
% \begin{align}
%     g_{2}(\textbf{r}_{12}) = \frac{\rho_{2}(\textbf{r}_{12})}{\rho^{2}},
%     %g_{2}(r) = \frac{\rho_{2}(r)}{\rho^{2}},    
% \end{align}
% where $\rho_{2}(\textbf{r}_{12})$ is the two-particle probability density function associated with finding two particles a distance $\textbf{r}_{12}$ apart, and $\rho$ is the number density.
% %at position $\textbf{r}_{1}$ and $\textbf{r}_{2}$
% The total correlation function is defined by 
% \begin{align}
%     h_{2}(\textbf{r}_{12}) = g_{2}(\textbf{r}_{12})-1.
%     %h(r) = g_{2}(r)-1.
% \end{align}

We consider a homogeneous point pattern in $d$-dimensional Euclidean space and an observation window $\Omega$ characterized by a length scale $R$ and a centroid $\textbf{x}_0$. In the case of spherical windows (circular in 2 dimensions), $R$ is simply the radius. We fix $\textbf{x}_0$ and study the relative fluctuations of the number of points in the observation window as $R$ grows large. We define a window indicator function  
\begin{align}\label{eq:WI}
w(\textbf{x}-\textbf{x}_{0}; R) = \begin{cases}    1, & \text{if $\textbf{x} \in \Omega$}.\\
    0, & \text{$\textbf{x} \notin \Omega$}.
  \end{cases}
\end{align}
The number of points located in a window is expressed by
\begin{align}\label{eq:Nwindow}
N(\textbf{x}_{0}; R) = \sum^{N}_{i=1} w(\textbf{r}_{i}-\textbf{x}_{0}; R)\, .
\end{align}
The average number of points contained within the window over many realizations is
\begin{align}\label{eq:Nwindow2}
\langle N(R) \rangle = \rho v(R),
\end{align}
where $\rho$ is the density and
$v(R) = \int_{\mathcal{R}^{d}} w(\textbf{r}; R) d\textbf{r}$ is the volume (area) of the window. In Eq.~\eqref{eq:Nwindow2}, we have dropped the dependence on $\mathbf{x}_0$ due to the assumption of heterogeneity. Similarly, by averaging the square of Eq.~\eqref{eq:Nwindow} we obtain 
\begin{align}\label{eq:Nvar}
\sigma^2_N(R) = \langle N(R) \rangle\Bigg[1 + \rho \int_{\mathcal{R}^{d}}\hspace{-0.2cm}h(\textbf{r})\alpha(\textbf{r};R)d\textbf{r}\Bigg],
\end{align}
where $h(\textbf{r})=\rho_2(\mathbf{r})/\rho^2-1$ and
\begin{align}\label{eq:alp}
\alpha(\textbf{r}; R) = 
\int_{\mathcal{R}^{d}}
\frac{w(\textbf{x}_{0}; R)w(\textbf{x}_{0} + \textbf{r}; R)}{v(R)}d\textbf{x}_{0} 
\end{align}
is the intersection volume of two windows whose centroids are separated by $\textbf{r}$, normalized by the volume of the window. We now move to Fourier space and use Parseval's theorem to rewrite Eq. \eqref{eq:Nvar} as
\begin{align}\label{eq:NvarFT}
\sigma^2_N(R)&=\langle N(R) \rangle\Bigg[1 + \frac{\rho}{(2 \pi)^{d}} \int_{\mathcal{R}^{d}}\hspace{-0.2cm}\tilde h(\textbf{k})\tilde\alpha(\textbf{k};R)d\textbf{k}\Bigg],
\end{align}
where we introduced the Fourier transform $\tilde{f}(\mathbf{k})=\int_{\mathcal{R}^{d}} f(\mathbf{r}) e^{-i\mathbf{k}\cdot\mathbf{r}}d\mathbf{r}$. 
We now note that the structure factor, defined in Eq.~\eqref{eq:structurefact}, is related with the Fourier transform of $h(\textbf{r})$ by
\begin{align}\label{eq:ft_SF}
    S(\textbf{k}) = 1 + \rho \tilde{h}(\textbf{k})\, .
\end{align}
% We now use that
% \begin{align}\label{eq:alpFT}
% \tilde\alpha(\textbf{k}; \textbf{R}) = \frac{\tilde w^{2}(\textbf{k}; \textbf{R})}{v(\textbf{R})},
% \end{align}

We use Eq.~\eqref{eq:ft_SF} to rewrite Eq. \eqref{eq:NvarFT} in terms of the structure factor as
\begin{align}\label{eq:nvdual}
 \sigma^2_N (R)= \langle N(R) \rangle \Bigg[\frac{1}{(2 \pi)^{d}}\hspace{-0.1cm} \int\hspace{-0cm}S(\textbf{k}) \tilde\alpha(\textbf{k};R) d\textbf{k}\Bigg]\, ,
\end{align}
where we also used that
\begin{align}\label{eq:alpFT1}
\frac{1}{(2 \pi)^{d}}\int_{\mathcal{R}^{d}}\tilde\alpha(\textbf{k}; R)d\textbf{k} = \alpha(\textbf{r}=\mathbf{0}; R) = 1.
\end{align}
We now consider the limit as the volume of the window tends to infinity. In this limit, the function $\alpha(\textbf{r};R)$ tends to one, so that $\tilde\alpha(\textbf{k};R)$ tends to $(2\pi)^{d}\delta(\textbf{k})$. We therefore obtain a relation between the number variance, the structure factor, and total correlation function:
\begin{align}
    \lim_{R \rightarrow \infty} \frac{\sigma^2_N(R)}{\langle N(R) \rangle} = 1+\rho \int_{\mathcal{R}^{d}}h(\textbf{r})d\textbf{r}=S(\textbf{k}=\mathbf{0}).
\end{align}
The behavior of this limit can be used to characterize three classes of point process:
\begin{itemize}
\item If the limit is finite, the point process is Poissonian-like.  In this case, the variance scales in the same way as the mean, $\alpha = 1$, and both the integral of $h(\textbf{r})$ and $S(\textbf{k}=\mathbf{0})$ are finite.
\item If the limit is zero, the point process is hyperuniform. In this case we have $\alpha < 1$, the integral of $h(\textbf{r})$ tends to $-1/\rho$, and  $S(\textbf{k})$ tends to zero for $\mathbf{k}\rightarrow \mathbf{0}$. The fact that the integral of $h(\textbf{r})$ is negative implies that the pair correlation function must be negative for certain values of $\textbf{r}$.  
\item If the limit diverges, the point process is hyperdisordered.
In this case, we have $\alpha > 1$, and the space integral of $h(\textbf{r})$ must diverge. In the Fourier space, the limit of  $S(\textbf{k}=\mathbf{0})$ for $\mathbf{k}\rightarrow \mathbf{0}$ must diverge as well.
\end{itemize}

\begin{figure*}[htb]
\includegraphics[width=0.9\textwidth]{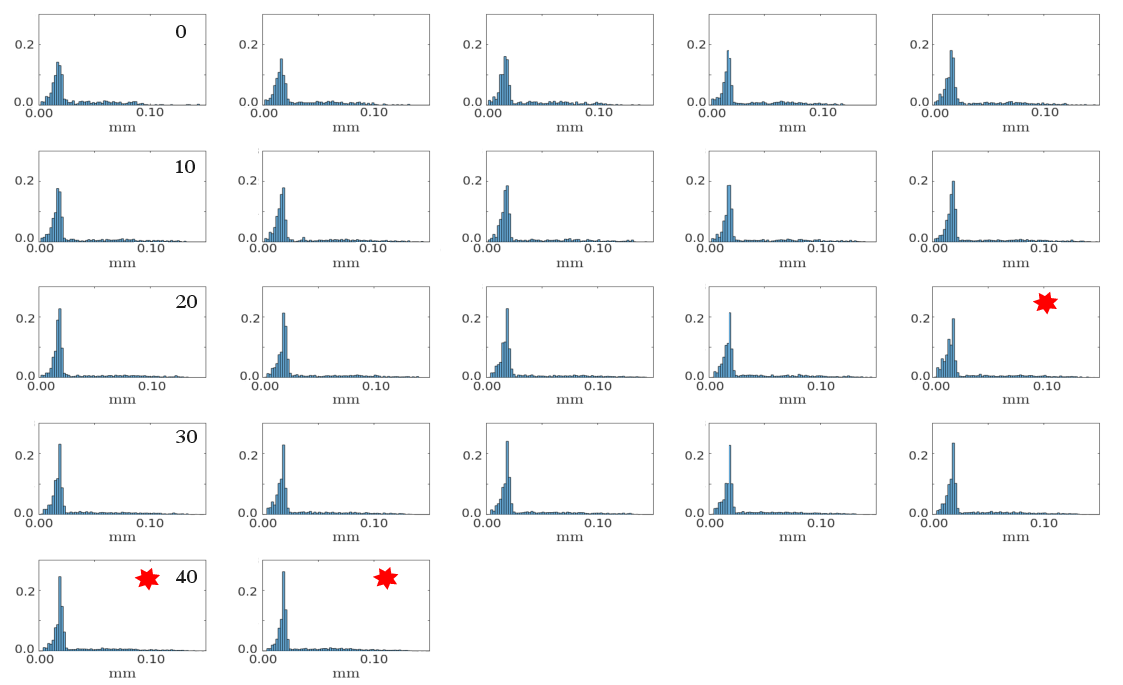}
\caption{Distribution of chromatophore radii for all 
experimental days. %$(0,2,4,6,8,10,12,14,16,18,20,22,24,26,28,30,32,34,40,42)$. 
%Days 36 and 38 are not shown to make the figure smaller. 
The red star indicates those distributions in which the null hypothesis was rejected according to the Kolmogorov-Smirnov test.}\label{fig:figA1}
\end{figure*}

\section{Iterative model}\label{app:iterative}

Here we derive the solution of the iterative model, Eq.~\eqref{eq:recurrence_formula}, and discuss why this expression implies that the model always presents hyperdisordered scaling. We start from with the structure factor of the initial condition
\begin{align}
    S_{0}(k) &= \frac{1}{n}\sum_{i,j}^{n}\langle e^{-ik(x_{i}-x_{j})} \rangle.
\end{align}
After the first iteration, we obtain
\begin{align}\label{eq2}
    S_{1}(k) &= \frac{1}{2n}\sum_{i,j}^{n}\Big(\langle e^{-2ik(x_{i}-x_{j})} \rangle + \langle e^{-2ik(x_{i}+\frac{\delta x_{i}}{2}-x_{j})} \rangle \nonumber\\
    &+ \langle e^{-2ik(x_{i}-x_{j}-\frac{\delta x_{j}}{2})} \rangle \nonumber \\ &+ \langle e^{-2ik(x_{i}-x_{j}+\frac{\delta x_{i}}{2}-\frac{\delta x_{j}}{2})} \rangle\Big),
\end{align}
where $\delta_{x}$ are the i.i.d. Gaussian random variables corresponding to the newly inserted points. We evaluate the expectations in Eq.~\eqref{eq2} using standard properties of Gaussian integrals, obtaining
\begin{align}
    S_{1}(k) &= \frac{S_{0}(2k)}{2}\Big[ 1 + \Big(e^{ik\Delta - \frac{k^{2}\sigma^{2}}{2}} + e^{-ik\Delta - \frac{k^{2}\sigma^{2}}{2}} \Big) + e^{-k^{2}\sigma^{2}}\Big] \nonumber\\
    &+ \frac{1}{2}(1 - e^{-k^2\sigma^2}),
\end{align}
which can be simplified to
\begin{align}
    S_{1}(k) &= \frac{S_{0}(2k)}{2}\Big[ 1 + 2 \mathrm{cos}(\Delta k)e^{-\frac{k^{2}\sigma^{2}}{2}} \nonumber\\
    &+ e^{-k^{2}\sigma^{2}}\Big] + \frac{1}{2}(1 - e^{-k^2\sigma^2}).
\end{align}
At the second iteration, we obtain
\begin{align}
S_{2}(k) &= \frac{1}{4n}\sum_{i,j}^{n}\Big(\langle e^{-4ik(x_{i}-x_{j})} \rangle + \langle e^{-4ik(x_{i}+\frac{\delta x_{i}}{2}-x_{j})} \rangle \nonumber\\
&+\langle e^{-4ik(x_{i}-x_{j}-\frac{\delta x_{j}}{2})} \rangle + \langle e^{-4ik(x_{i}-x_{j}+\frac{\delta x_{i}}{2}-\frac{\delta x_{j}}{2})} \rangle\Big)\nonumber\\
&\times\Big[ 1 + 2 \mathrm{cos}(\Delta k)e^{-\frac{k^{2}\sigma^{2}}{2}} + e^{-k^{2}\sigma^{2}}\Big] + \frac{1}{2}(1 - e^{-k^2\sigma^2}),
\end{align}
which again can be simplified to
\begin{align}
S_{2}(k) &= \Bigg(\frac{S_{0}(4k)}{4}\Big[ 1 + 2 \mathrm{cos}(2\Delta k)e^{-k^{2}\sigma^{2}} + e^{-2k^{2}\sigma^{2}}\Big] \nonumber \\ &+ \frac{1}{2}(1 - e^{-2k^2\sigma^2})\Bigg)\Big[ 1 + 2 \mathrm{cos}(\Delta k)e^{-\frac{k^{2}\sigma^{2}}{2}} + e^{-k^{2}\sigma^{2}}\Big] \nonumber\\ &+ \frac{1}{2}(1 - e^{-k^2\sigma^2}).
\end{align}
By iterating this procedure, we find that the structure factor at the $t$-th iteration is expressed by
Eq.~\eqref{eq:recurrence_formula}. 

% We now want to show that the structure factor diverges for $k\rightarrow 0$ for an infinite system ($t\rightarrow \infty$). We set $k = \left(2 \pi /(n 2^{t} \right))$ in Eq. \eqref{eq:recurrence_formula}, where $n 2^{t}$ is the number of cells after $t$ iterations. We then obtain
% \begin{equation}
% \lim_{t\rightarrow\infty}S_{t}(2\pi/(n 2^{t})) \rightarrow \infty.
% \end{equation}
% This divergence implies a hyperdisordered behavior. The mathematical reason for this divergence is that, for some $i < t$, we have that 
% \begin{align}\label{eq:Flimit}
% F\left(\frac{2^{t-i} 2 \pi}{n 2^{t}}\right) &= 1 + 2\cos\left(\frac{\Delta 2 \pi}{n 2^{t}}\right)e^{-\frac{1}{2}\left(\frac{2^{t-i} 2\pi \sigma}{n 2^{t}}\right)^{2}} \nonumber\\
% &+ e^{-\left(\frac{2^{t-i} 2\pi \sigma}{n 2^{t}}\right)^{2}} > 2.
% \end{align}

\section{Kolmogorov-Smirnov test for comparing chromatophore radius distribution through time}\label{app:KS}

We pooled chromatophore radii from 3 squids and binned into equally-sized bins $(0,0.01,0.02,...,0.18)$ for each $T_{\expp} \in (0,2,4,...,42)$. We then performed two separate analyses using these distributions: (1) each $T_{\expp}$ was compared to the distribution made by aggregating chromatophore radii data across all times, (2) each $T_{\expp}$ was compared individually against all other $T_{\expp}$.  In the case of (1), from 22 tests the null hypothesis was rejected 3 times at $\beta = 0.05$.  In the case of (2), from 231 tests the null hypothesis was rejected 27 times, also at $\beta = 0.05$. This means that, for both analyses the null hypothesis was not rejected in approximately $85\%$ of cases, supporting the conclusion that the chromatophore size distribution is stationary.  These distributions are shown in Fig.~\ref{fig:figA1}.  Those in which the null hypothesis was rejected are indicated by a red star.

\section{Inference of chromatophore insertion times}\label{app:insertiontimes}

We pooled new chromatophores for each $T_{\expp} \in (0,2,4,...,42)$ across 3 squids to generate the chromatophore trajectories in Fig. \ref{fig6}b.  $T_{\squid}$ at $T_{\exp} = 0$ was determined by minimising the residuals between chromatophore trajectories according to the function $T_\squid/(T_{\squid} - t_{\chr})$.  This resulted in $T_{\squid} = 30.7$ days. The corresponding trajectories are displayed in Fig. \ref{fig6}c. From these trajectories, we generated a master curve using MATLAB's curve-fitting tool box (see below for function) and minimized the residuals of remaining chromatophores whose size coincided with this master curve using an interpolation scheme \cite{bhattacharjee2001measure}. Following this procedure, we generated a new master curve, and minimized  the residuals of remaining chromatophores whose size coincided with this master curve. We iterated this process until all remaining chromatophores had been matched.  The final master curve is displayed in Fig. 6c.

We fitted the function describing the growth of the chromatophore radius with respect to $\tau$ to a $5^{th}$ degree rational using MATLAB's curve-fitting toolbox.  
The equation is:
\begin{equation}\label{eq:cf}
R(\tau) =  \frac{(p_1\tau^5 + p_2\tau^4 + p_3\tau^3 + p_4\tau^2 + p_5\tau^1 + p_6)}{(\tau^5 + q_1\tau^4 + q_2\tau^3 + q_3\tau^2 + q_4\tau^1 + q_5)},
\end{equation}
and the initial size of a chromatophore $R(1) = c_{r} = 1.25 \cdot 10^{-2}$. The coefficient values are given in Table \ref{tab:table1}.
\begin{table}[h!]
\caption{Coefficient values}
\centering
\begin{tabular}{ll}
$p_1$ & 0.15 \\
$p_2$ & -2.21 \\
$p_3$ & 11.95 \\
$p_4$ & -27.98 \\
$p_5$ & 28.77 \\
$p_6$ & -10.25 \\
$q_1$ & -13.09 \\
$q_2$ & 56.39 \\
$q_3$ & -67.27 \\
$q_4$ & -50.63 \\
$q_5$ & 108.4
\end{tabular}
\label{tab:table1}
\end{table}
The support for this function is $\tau\in[1, 7.55]$. For $\tau > 7.55$, we implement a linear function 
$$R(\tau) = c_{r2} + 2.42\cdot 10^{-4}\tau,$$
where $c_{r2} = R(7.55) = 0.126$.

\section{Data and code availability}

Data are available from the corresponding authors on request. Simulation and analysis code will be made available at \texttt{https://github.com/oist/hyperdisordered\_squid}. 

\bibliography{squidpattern}

\begin{thebibliography}{63}%
\makeatletter
\providecommand \@ifxundefined [1]{%
 \@ifx{#1\undefined}
}%
\providecommand \@ifnum [1]{%
 \ifnum #1\expandafter \@firstoftwo
 \else \expandafter \@secondoftwo
 \fi
}%
\providecommand \@ifx [1]{%
 \ifx #1\expandafter \@firstoftwo
 \else \expandafter \@secondoftwo
 \fi
}%
\providecommand \natexlab [1]{#1}%
\providecommand \enquote  [1]{``#1''}%
\providecommand \bibnamefont  [1]{#1}%
\providecommand \bibfnamefont [1]{#1}%
\providecommand \citenamefont [1]{#1}%
\providecommand \href@noop [0]{\@secondoftwo}%
\providecommand \href [0]{\begingroup \@sanitize@url \@href}%
\providecommand \@href[1]{\@@startlink{#1}\@@href}%
\providecommand \@@href[1]{\endgroup#1\@@endlink}%
\providecommand \@sanitize@url [0]{\catcode `\\12\catcode `\$12\catcode
  `\&12\catcode `\#12\catcode `\^12\catcode `\_12\catcode `\%12\relax}%
\providecommand \@@startlink[1]{}%
\providecommand \@@endlink[0]{}%
\providecommand \url  [0]{\begingroup\@sanitize@url \@url }%
\providecommand \@url [1]{\endgroup\@href {#1}{\urlprefix }}%
\providecommand \urlprefix  [0]{URL }%
\providecommand \Eprint [0]{\href }%
\providecommand \doibase [0]{https://doi.org/}%
\providecommand \selectlanguage [0]{\@gobble}%
\providecommand \bibinfo  [0]{\@secondoftwo}%
\providecommand \bibfield  [0]{\@secondoftwo}%
\providecommand \translation [1]{[#1]}%
\providecommand \BibitemOpen [0]{}%
\providecommand \bibitemStop [0]{}%
\providecommand \bibitemNoStop [0]{.\EOS\space}%
\providecommand \EOS [0]{\spacefactor3000\relax}%
\providecommand \BibitemShut  [1]{\csname bibitem#1\endcsname}%
\let\auto@bib@innerbib\@empty
%</preamble>
\bibitem [{\citenamefont {Hansen}\ and\ \citenamefont
  {McDonald}(2013)}]{hansen2013theory}%
  \BibitemOpen
  \bibfield  {author} {\bibinfo {author} {\bibfnamefont {J.-P.}\ \bibnamefont
  {Hansen}}\ and\ \bibinfo {author} {\bibfnamefont {I.~R.}\ \bibnamefont
  {McDonald}},\ }\href@noop {} {\emph {\bibinfo {title} {Theory of simple
  liquids: with applications to soft matter}}}\ (\bibinfo  {publisher}
  {Academic press},\ \bibinfo {year} {2013})\BibitemShut {NoStop}%
\bibitem [{\citenamefont {Parisi}\ \emph {et~al.}(2020)\citenamefont {Parisi},
  \citenamefont {Urbani},\ and\ \citenamefont {Zamponi}}]{parisi2020theory}%
  \BibitemOpen
  \bibfield  {author} {\bibinfo {author} {\bibfnamefont {G.}~\bibnamefont
  {Parisi}}, \bibinfo {author} {\bibfnamefont {P.}~\bibnamefont {Urbani}},\
  and\ \bibinfo {author} {\bibfnamefont {F.}~\bibnamefont {Zamponi}},\
  }\href@noop {} {\emph {\bibinfo {title} {Theory of simple glasses: exact
  solutions in infinite dimensions}}}\ (\bibinfo  {publisher} {Cambridge
  University Press},\ \bibinfo {year} {2020})\BibitemShut {NoStop}%
\bibitem [{\citenamefont {Duran}(2012)}]{duran2012sands}%
  \BibitemOpen
  \bibfield  {author} {\bibinfo {author} {\bibfnamefont {J.}~\bibnamefont
  {Duran}},\ }\href@noop {} {\emph {\bibinfo {title} {Sands, powders, and
  grains: an introduction to the physics of granular materials}}}\ (\bibinfo
  {publisher} {Springer Science \& Business Media},\ \bibinfo {year}
  {2012})\BibitemShut {NoStop}%
\bibitem [{\citenamefont {R\'enyi}(1958)}]{renyi1963}%
  \BibitemOpen
  \bibfield  {author} {\bibinfo {author} {\bibfnamefont {A.}~\bibnamefont
  {R\'enyi}},\ }\bibfield  {title} {\bibinfo {title} {On a one-dimensional
  problem concerning random space-filling},\ }\href@noop {} {\bibfield
  {journal} {\bibinfo  {journal} {Publ. Math. Inst. Hung. Acad. Sci.}\ }\textbf
  {\bibinfo {volume} {3}},\ \bibinfo {pages} {109} (\bibinfo {year}
  {1958})}\BibitemShut {NoStop}%
\bibitem [{\citenamefont {Donev}\ \emph {et~al.}(2004)\citenamefont {Donev},
  \citenamefont {Cisse}, \citenamefont {Sachs}, \citenamefont {Variano},
  \citenamefont {Stillinger}, \citenamefont {Connelly}, \citenamefont
  {Torquato},\ and\ \citenamefont {Chaikin}}]{donev2004improving}%
  \BibitemOpen
  \bibfield  {author} {\bibinfo {author} {\bibfnamefont {A.}~\bibnamefont
  {Donev}}, \bibinfo {author} {\bibfnamefont {I.}~\bibnamefont {Cisse}},
  \bibinfo {author} {\bibfnamefont {D.}~\bibnamefont {Sachs}}, \bibinfo
  {author} {\bibfnamefont {E.~A.}\ \bibnamefont {Variano}}, \bibinfo {author}
  {\bibfnamefont {F.~H.}\ \bibnamefont {Stillinger}}, \bibinfo {author}
  {\bibfnamefont {R.}~\bibnamefont {Connelly}}, \bibinfo {author}
  {\bibfnamefont {S.}~\bibnamefont {Torquato}},\ and\ \bibinfo {author}
  {\bibfnamefont {P.~M.}\ \bibnamefont {Chaikin}},\ }\bibfield  {title}
  {\bibinfo {title} {Improving the density of jammed disordered packings using
  ellipsoids},\ }\href@noop {} {\bibfield  {journal} {\bibinfo  {journal}
  {Science}\ }\textbf {\bibinfo {volume} {303}},\ \bibinfo {pages} {990}
  (\bibinfo {year} {2004})}\BibitemShut {NoStop}%
\bibitem [{\citenamefont {Torquato}(2018)}]{torquato2018hyperuniform}%
  \BibitemOpen
  \bibfield  {author} {\bibinfo {author} {\bibfnamefont {S.}~\bibnamefont
  {Torquato}},\ }\bibfield  {title} {\bibinfo {title} {Hyperuniform states of
  matter},\ }\href@noop {} {\bibfield  {journal} {\bibinfo  {journal} {Physics
  Reports}\ }\textbf {\bibinfo {volume} {745}},\ \bibinfo {pages} {1} (\bibinfo
  {year} {2018})}\BibitemShut {NoStop}%
\bibitem [{\citenamefont {You}\ \emph {et~al.}(2021)\citenamefont {You},
  \citenamefont {Pearce},\ and\ \citenamefont {Giomi}}]{you2021confinement}%
  \BibitemOpen
  \bibfield  {author} {\bibinfo {author} {\bibfnamefont {Z.}~\bibnamefont
  {You}}, \bibinfo {author} {\bibfnamefont {D.~J.}\ \bibnamefont {Pearce}},\
  and\ \bibinfo {author} {\bibfnamefont {L.}~\bibnamefont {Giomi}},\ }\bibfield
   {title} {\bibinfo {title} {Confinement-induced self-organization in growing
  bacterial colonies},\ }\href@noop {} {\bibfield  {journal} {\bibinfo
  {journal} {Science Advances}\ }\textbf {\bibinfo {volume} {7}},\ \bibinfo
  {pages} {eabc8685} (\bibinfo {year} {2021})}\BibitemShut {NoStop}%
\bibitem [{\citenamefont {Cheng}\ \emph {et~al.}(2014)\citenamefont {Cheng},
  \citenamefont {Niu}, \citenamefont {Warren}, \citenamefont {Pevny},
  \citenamefont {Lovell-Badge}, \citenamefont {Hwa},\ and\ \citenamefont
  {Cheah}}]{cheng2014predicting}%
  \BibitemOpen
  \bibfield  {author} {\bibinfo {author} {\bibfnamefont {C.~W.}\ \bibnamefont
  {Cheng}}, \bibinfo {author} {\bibfnamefont {B.}~\bibnamefont {Niu}}, \bibinfo
  {author} {\bibfnamefont {M.}~\bibnamefont {Warren}}, \bibinfo {author}
  {\bibfnamefont {L.~H.}\ \bibnamefont {Pevny}}, \bibinfo {author}
  {\bibfnamefont {R.}~\bibnamefont {Lovell-Badge}}, \bibinfo {author}
  {\bibfnamefont {T.}~\bibnamefont {Hwa}},\ and\ \bibinfo {author}
  {\bibfnamefont {K.~S.}\ \bibnamefont {Cheah}},\ }\bibfield  {title} {\bibinfo
  {title} {Predicting the spatiotemporal dynamics of hair follicle patterns in
  the developing mouse},\ }\href@noop {} {\bibfield  {journal} {\bibinfo
  {journal} {Proceedings of the National Academy of Sciences}\ }\textbf
  {\bibinfo {volume} {111}},\ \bibinfo {pages} {2596} (\bibinfo {year}
  {2014})}\BibitemShut {NoStop}%
\bibitem [{\citenamefont {Shyer}\ \emph {et~al.}(2017)\citenamefont {Shyer},
  \citenamefont {Rodrigues}, \citenamefont {Schroeder}, \citenamefont
  {Kassianidou}, \citenamefont {Kumar},\ and\ \citenamefont
  {Harland}}]{shyer2017emergent}%
  \BibitemOpen
  \bibfield  {author} {\bibinfo {author} {\bibfnamefont {A.~E.}\ \bibnamefont
  {Shyer}}, \bibinfo {author} {\bibfnamefont {A.~R.}\ \bibnamefont
  {Rodrigues}}, \bibinfo {author} {\bibfnamefont {G.~G.}\ \bibnamefont
  {Schroeder}}, \bibinfo {author} {\bibfnamefont {E.}~\bibnamefont
  {Kassianidou}}, \bibinfo {author} {\bibfnamefont {S.}~\bibnamefont {Kumar}},\
  and\ \bibinfo {author} {\bibfnamefont {R.~M.}\ \bibnamefont {Harland}},\
  }\bibfield  {title} {\bibinfo {title} {Emergent cellular self-organization
  and mechanosensation initiate follicle pattern in the avian skin},\
  }\href@noop {} {\bibfield  {journal} {\bibinfo  {journal} {Science}\ }\textbf
  {\bibinfo {volume} {357}},\ \bibinfo {pages} {811} (\bibinfo {year}
  {2017})}\BibitemShut {NoStop}%
\bibitem [{\citenamefont {Mongera}\ \emph {et~al.}(2018)\citenamefont
  {Mongera}, \citenamefont {Rowghanian}, \citenamefont {Gustafson},
  \citenamefont {Shelton}, \citenamefont {Kealhofer}, \citenamefont {Carn},
  \citenamefont {Serwane}, \citenamefont {Lucio}, \citenamefont {Giammona},\
  and\ \citenamefont {Camp{\`a}s}}]{mongera2018fluid}%
  \BibitemOpen
  \bibfield  {author} {\bibinfo {author} {\bibfnamefont {A.}~\bibnamefont
  {Mongera}}, \bibinfo {author} {\bibfnamefont {P.}~\bibnamefont {Rowghanian}},
  \bibinfo {author} {\bibfnamefont {H.~J.}\ \bibnamefont {Gustafson}}, \bibinfo
  {author} {\bibfnamefont {E.}~\bibnamefont {Shelton}}, \bibinfo {author}
  {\bibfnamefont {D.~A.}\ \bibnamefont {Kealhofer}}, \bibinfo {author}
  {\bibfnamefont {E.~K.}\ \bibnamefont {Carn}}, \bibinfo {author}
  {\bibfnamefont {F.}~\bibnamefont {Serwane}}, \bibinfo {author} {\bibfnamefont
  {A.~A.}\ \bibnamefont {Lucio}}, \bibinfo {author} {\bibfnamefont
  {J.}~\bibnamefont {Giammona}},\ and\ \bibinfo {author} {\bibfnamefont
  {O.}~\bibnamefont {Camp{\`a}s}},\ }\bibfield  {title} {\bibinfo {title} {A
  fluid-to-solid jamming transition underlies vertebrate body axis
  elongation},\ }\href@noop {} {\bibfield  {journal} {\bibinfo  {journal}
  {Nature}\ }\textbf {\bibinfo {volume} {561}},\ \bibinfo {pages} {401}
  (\bibinfo {year} {2018})}\BibitemShut {NoStop}%
\bibitem [{\citenamefont {Gottheil}\ \emph {et~al.}(2023)\citenamefont
  {Gottheil}, \citenamefont {Lippoldt}, \citenamefont {Grosser}, \citenamefont
  {Renner}, \citenamefont {Saibah}, \citenamefont {Tschodu}, \citenamefont
  {Po{\ss}{\"o}gel}, \citenamefont {Wegscheider}, \citenamefont {Ulm},
  \citenamefont {Friedrichs} \emph {et~al.}}]{gottheil2023state}%
  \BibitemOpen
  \bibfield  {author} {\bibinfo {author} {\bibfnamefont {P.}~\bibnamefont
  {Gottheil}}, \bibinfo {author} {\bibfnamefont {J.}~\bibnamefont {Lippoldt}},
  \bibinfo {author} {\bibfnamefont {S.}~\bibnamefont {Grosser}}, \bibinfo
  {author} {\bibfnamefont {F.}~\bibnamefont {Renner}}, \bibinfo {author}
  {\bibfnamefont {M.}~\bibnamefont {Saibah}}, \bibinfo {author} {\bibfnamefont
  {D.}~\bibnamefont {Tschodu}}, \bibinfo {author} {\bibfnamefont {A.-K.}\
  \bibnamefont {Po{\ss}{\"o}gel}}, \bibinfo {author} {\bibfnamefont {A.-S.}\
  \bibnamefont {Wegscheider}}, \bibinfo {author} {\bibfnamefont
  {B.}~\bibnamefont {Ulm}}, \bibinfo {author} {\bibfnamefont {K.}~\bibnamefont
  {Friedrichs}}, \emph {et~al.},\ }\bibfield  {title} {\bibinfo {title} {State
  of cell unjamming correlates with distant metastasis in cancer patients},\
  }\href@noop {} {\bibfield  {journal} {\bibinfo  {journal} {Physical Review
  X}\ }\textbf {\bibinfo {volume} {13}},\ \bibinfo {pages} {031003} (\bibinfo
  {year} {2023})}\BibitemShut {NoStop}%
\bibitem [{\citenamefont {Torquato}\ and\ \citenamefont
  {Stillinger}(2003)}]{torquato2003local}%
  \BibitemOpen
  \bibfield  {author} {\bibinfo {author} {\bibfnamefont {S.}~\bibnamefont
  {Torquato}}\ and\ \bibinfo {author} {\bibfnamefont {F.~H.}\ \bibnamefont
  {Stillinger}},\ }\bibfield  {title} {\bibinfo {title} {Local density
  fluctuations, hyperuniformity, and order metrics},\ }\href@noop {} {\bibfield
   {journal} {\bibinfo  {journal} {Physical Review E}\ }\textbf {\bibinfo
  {volume} {68}},\ \bibinfo {pages} {041113} (\bibinfo {year}
  {2003})}\BibitemShut {NoStop}%
\bibitem [{\citenamefont {Zachary}\ \emph {et~al.}(2011)\citenamefont
  {Zachary}, \citenamefont {Jiao},\ and\ \citenamefont
  {Torquato}}]{zachary2011hyperuniformity}%
  \BibitemOpen
  \bibfield  {author} {\bibinfo {author} {\bibfnamefont {C.~E.}\ \bibnamefont
  {Zachary}}, \bibinfo {author} {\bibfnamefont {Y.}~\bibnamefont {Jiao}},\ and\
  \bibinfo {author} {\bibfnamefont {S.}~\bibnamefont {Torquato}},\ }\bibfield
  {title} {\bibinfo {title} {Hyperuniformity, quasi-long-range correlations,
  and void-space constraints in maximally random jammed particle packings. i.
  polydisperse spheres},\ }\href@noop {} {\bibfield  {journal} {\bibinfo
  {journal} {Physical Review E}\ }\textbf {\bibinfo {volume} {83}},\ \bibinfo
  {pages} {051308} (\bibinfo {year} {2011})}\BibitemShut {NoStop}%
\bibitem [{\citenamefont {Dreyfus}\ \emph {et~al.}(2015)\citenamefont
  {Dreyfus}, \citenamefont {Xu}, \citenamefont {Still}, \citenamefont {Hough},
  \citenamefont {Yodh},\ and\ \citenamefont
  {Torquato}}]{dreyfus2015diagnosing}%
  \BibitemOpen
  \bibfield  {author} {\bibinfo {author} {\bibfnamefont {R.}~\bibnamefont
  {Dreyfus}}, \bibinfo {author} {\bibfnamefont {Y.}~\bibnamefont {Xu}},
  \bibinfo {author} {\bibfnamefont {T.}~\bibnamefont {Still}}, \bibinfo
  {author} {\bibfnamefont {L.~A.}\ \bibnamefont {Hough}}, \bibinfo {author}
  {\bibfnamefont {A.}~\bibnamefont {Yodh}},\ and\ \bibinfo {author}
  {\bibfnamefont {S.}~\bibnamefont {Torquato}},\ }\bibfield  {title} {\bibinfo
  {title} {Diagnosing hyperuniformity in two-dimensional, disordered, jammed
  packings of soft spheres},\ }\href@noop {} {\bibfield  {journal} {\bibinfo
  {journal} {Physical Review E}\ }\textbf {\bibinfo {volume} {91}},\ \bibinfo
  {pages} {012302} (\bibinfo {year} {2015})}\BibitemShut {NoStop}%
\bibitem [{\citenamefont {Wiese}(2024)}]{wiese2024hyperuniformity}%
  \BibitemOpen
  \bibfield  {author} {\bibinfo {author} {\bibfnamefont {K.~J.}\ \bibnamefont
  {Wiese}},\ }\bibfield  {title} {\bibinfo {title} {Hyperuniformity in the
  manna model, conserved directed percolation and depinning},\ }\href@noop {}
  {\bibfield  {journal} {\bibinfo  {journal} {Physical Review Letters}\
  }\textbf {\bibinfo {volume} {133}},\ \bibinfo {pages} {067103} (\bibinfo
  {year} {2024})}\BibitemShut {NoStop}%
\bibitem [{\citenamefont {De~Luca}\ \emph {et~al.}(2024)\citenamefont
  {De~Luca}, \citenamefont {Ma}, \citenamefont {Nardini},\ and\ \citenamefont
  {Cates}}]{de2024hyperuniformity}%
  \BibitemOpen
  \bibfield  {author} {\bibinfo {author} {\bibfnamefont {F.}~\bibnamefont
  {De~Luca}}, \bibinfo {author} {\bibfnamefont {X.}~\bibnamefont {Ma}},
  \bibinfo {author} {\bibfnamefont {C.}~\bibnamefont {Nardini}},\ and\ \bibinfo
  {author} {\bibfnamefont {M.~E.}\ \bibnamefont {Cates}},\ }\bibfield  {title}
  {\bibinfo {title} {Hyperuniformity in phase ordering: the roles of activity,
  noise, and non-constant mobility},\ }\href@noop {} {\bibfield  {journal}
  {\bibinfo  {journal} {J. Phys.: Condens. Matter}\ }\textbf {\bibinfo {volume}
  {36}},\ \bibinfo {pages} {405101} (\bibinfo {year} {2024})}\BibitemShut
  {NoStop}%
\bibitem [{\citenamefont {Backofen}\ \emph {et~al.}(2024)\citenamefont
  {Backofen}, \citenamefont {Altawil}, \citenamefont {Salvalaglio},\ and\
  \citenamefont {Voigt}}]{backofen2024nonequilibrium}%
  \BibitemOpen
  \bibfield  {author} {\bibinfo {author} {\bibfnamefont {R.}~\bibnamefont
  {Backofen}}, \bibinfo {author} {\bibfnamefont {A.~Y.}\ \bibnamefont
  {Altawil}}, \bibinfo {author} {\bibfnamefont {M.}~\bibnamefont
  {Salvalaglio}},\ and\ \bibinfo {author} {\bibfnamefont {A.}~\bibnamefont
  {Voigt}},\ }\bibfield  {title} {\bibinfo {title} {Nonequilibrium hyperuniform
  states in active turbulence},\ }\href@noop {} {\bibfield  {journal} {\bibinfo
   {journal} {Proceedings of the National Academy of Sciences}\ }\textbf
  {\bibinfo {volume} {121}},\ \bibinfo {pages} {e2320719121} (\bibinfo {year}
  {2024})}\BibitemShut {NoStop}%
\bibitem [{\citenamefont {Lei}\ and\ \citenamefont {Ni}(2024)}]{lei2024non}%
  \BibitemOpen
  \bibfield  {author} {\bibinfo {author} {\bibfnamefont {Y.}~\bibnamefont
  {Lei}}\ and\ \bibinfo {author} {\bibfnamefont {R.}~\bibnamefont {Ni}},\
  }\bibfield  {title} {\bibinfo {title} {Non-equilibrium dynamic hyperuniform
  states},\ }\href@noop {} {\bibfield  {journal} {\bibinfo  {journal} {arXiv
  preprint arXiv:2405.12818}\ } (\bibinfo {year} {2024})}\BibitemShut {NoStop}%
\bibitem [{\citenamefont {Jiao}\ \emph {et~al.}(2014)\citenamefont {Jiao},
  \citenamefont {Lau}, \citenamefont {Hatzikirou}, \citenamefont
  {Meyer-Hermann}, \citenamefont {Corbo},\ and\ \citenamefont
  {Torquato}}]{jiao2014avian}%
  \BibitemOpen
  \bibfield  {author} {\bibinfo {author} {\bibfnamefont {Y.}~\bibnamefont
  {Jiao}}, \bibinfo {author} {\bibfnamefont {T.}~\bibnamefont {Lau}}, \bibinfo
  {author} {\bibfnamefont {H.}~\bibnamefont {Hatzikirou}}, \bibinfo {author}
  {\bibfnamefont {M.}~\bibnamefont {Meyer-Hermann}}, \bibinfo {author}
  {\bibfnamefont {J.~C.}\ \bibnamefont {Corbo}},\ and\ \bibinfo {author}
  {\bibfnamefont {S.}~\bibnamefont {Torquato}},\ }\bibfield  {title} {\bibinfo
  {title} {Avian photoreceptor patterns represent a disordered hyperuniform
  solution to a multiscale packing problem},\ }\href@noop {} {\bibfield
  {journal} {\bibinfo  {journal} {Physical Review E}\ }\textbf {\bibinfo
  {volume} {89}},\ \bibinfo {pages} {022721} (\bibinfo {year}
  {2014})}\BibitemShut {NoStop}%
\bibitem [{\citenamefont {Liu}\ \emph {et~al.}(2024)\citenamefont {Liu},
  \citenamefont {Chen}, \citenamefont {Tian}, \citenamefont {Xu},\ and\
  \citenamefont {Jiao}}]{liu2024universal}%
  \BibitemOpen
  \bibfield  {author} {\bibinfo {author} {\bibfnamefont {Y.}~\bibnamefont
  {Liu}}, \bibinfo {author} {\bibfnamefont {D.}~\bibnamefont {Chen}}, \bibinfo
  {author} {\bibfnamefont {J.}~\bibnamefont {Tian}}, \bibinfo {author}
  {\bibfnamefont {W.}~\bibnamefont {Xu}},\ and\ \bibinfo {author}
  {\bibfnamefont {Y.}~\bibnamefont {Jiao}},\ }\bibfield  {title} {\bibinfo
  {title} {Universal hyperuniform organization in looped leaf vein networks},\
  }\href@noop {} {\bibfield  {journal} {\bibinfo  {journal} {Physical Review
  Letters}\ }\textbf {\bibinfo {volume} {133}},\ \bibinfo {pages} {028401}
  (\bibinfo {year} {2024})}\BibitemShut {NoStop}%
\bibitem [{\citenamefont {Torquato}(2021)}]{Torquato2021}%
  \BibitemOpen
  \bibfield  {author} {\bibinfo {author} {\bibfnamefont {S.}~\bibnamefont
  {Torquato}},\ }\bibfield  {title} {\bibinfo {title} {Structural
  characterization of many-particle systems on approach to hyperuniform
  states},\ }\href {https://doi.org/10.1103/PhysRevE.103.052126} {\bibfield
  {journal} {\bibinfo  {journal} {Phys. Rev. E}\ }\textbf {\bibinfo {volume}
  {103}},\ \bibinfo {pages} {052126} (\bibinfo {year} {2021})}\BibitemShut
  {NoStop}%
\bibitem [{\citenamefont {Maher}\ and\ \citenamefont
  {Torquato}(2024{\natexlab{a}})}]{Emmett2024}%
  \BibitemOpen
  \bibfield  {author} {\bibinfo {author} {\bibfnamefont {C.~E.}\ \bibnamefont
  {Maher}}\ and\ \bibinfo {author} {\bibfnamefont {S.}~\bibnamefont
  {Torquato}},\ }\bibfield  {title} {\bibinfo {title} {Local order metrics for
  many-particle systems across length scales},\ }\href
  {https://doi.org/10.1103/PhysRevResearch.6.033262} {\bibfield  {journal}
  {\bibinfo  {journal} {Phys. Rev. Res.}\ }\textbf {\bibinfo {volume} {6}},\
  \bibinfo {pages} {033262} (\bibinfo {year} {2024}{\natexlab{a}})}\BibitemShut
  {NoStop}%
\bibitem [{\citenamefont {Gabrielli}\ \emph {et~al.}(2005)\citenamefont
  {Gabrielli}, \citenamefont {Labina}, \citenamefont {Joyce},\ and\
  \citenamefont {Pietronero}}]{gabrielli2005statistical}%
  \BibitemOpen
  \bibfield  {author} {\bibinfo {author} {\bibfnamefont {A.}~\bibnamefont
  {Gabrielli}}, \bibinfo {author} {\bibfnamefont {F.~S.}\ \bibnamefont
  {Labina}}, \bibinfo {author} {\bibfnamefont {M.}~\bibnamefont {Joyce}},\ and\
  \bibinfo {author} {\bibfnamefont {L.}~\bibnamefont {Pietronero}},\
  }\href@noop {} {\emph {\bibinfo {title} {Statistical physics for cosmic
  structures}}},\ Vol.\ \bibinfo {volume} {629}\ (\bibinfo  {publisher}
  {Springer Science \& Business Media},\ \bibinfo {year} {2005})\BibitemShut
  {NoStop}%
\bibitem [{\citenamefont {Palacci}\ \emph {et~al.}(2013)\citenamefont
  {Palacci}, \citenamefont {Sacanna}, \citenamefont {Steinberg}, \citenamefont
  {Pine},\ and\ \citenamefont {Chaikin}}]{palacci2013living}%
  \BibitemOpen
  \bibfield  {author} {\bibinfo {author} {\bibfnamefont {J.}~\bibnamefont
  {Palacci}}, \bibinfo {author} {\bibfnamefont {S.}~\bibnamefont {Sacanna}},
  \bibinfo {author} {\bibfnamefont {A.~P.}\ \bibnamefont {Steinberg}}, \bibinfo
  {author} {\bibfnamefont {D.~J.}\ \bibnamefont {Pine}},\ and\ \bibinfo
  {author} {\bibfnamefont {P.~M.}\ \bibnamefont {Chaikin}},\ }\bibfield
  {title} {\bibinfo {title} {Living crystals of light-activated colloidal
  surfers},\ }\href@noop {} {\bibfield  {journal} {\bibinfo  {journal}
  {Science}\ }\textbf {\bibinfo {volume} {339}},\ \bibinfo {pages} {936}
  (\bibinfo {year} {2013})}\BibitemShut {NoStop}%
\bibitem [{\citenamefont {Deseigne}\ \emph {et~al.}(2010)\citenamefont
  {Deseigne}, \citenamefont {Dauchot},\ and\ \citenamefont
  {Chat\'e}}]{Deseigne2010}%
  \BibitemOpen
  \bibfield  {author} {\bibinfo {author} {\bibfnamefont {J.}~\bibnamefont
  {Deseigne}}, \bibinfo {author} {\bibfnamefont {O.}~\bibnamefont {Dauchot}},\
  and\ \bibinfo {author} {\bibfnamefont {H.}~\bibnamefont {Chat\'e}},\
  }\bibfield  {title} {\bibinfo {title} {Collective motion of vibrated polar
  disks},\ }\href {https://doi.org/10.1103/PhysRevLett.105.098001} {\bibfield
  {journal} {\bibinfo  {journal} {Phys. Rev. Lett.}\ }\textbf {\bibinfo
  {volume} {105}},\ \bibinfo {pages} {098001} (\bibinfo {year}
  {2010})}\BibitemShut {NoStop}%
\bibitem [{\citenamefont {Puig}\ \emph {et~al.}(2024)\citenamefont {Puig},
  \citenamefont {S\'anchez}, \citenamefont {Herrera}, \citenamefont
  {Guillam\'on}, \citenamefont {Pribulov\'a}, \citenamefont {Ka\ifmmode
  \check{c}\else \v{c}\fi{}mar\ifmmode~\check{c}\else \v{c}\fi{}\'{\i}k},
  \citenamefont {Suderow}, \citenamefont {Kolton},\ and\ \citenamefont
  {Fasano}}]{Puig2024}%
  \BibitemOpen
  \bibfield  {author} {\bibinfo {author} {\bibfnamefont {J.}~\bibnamefont
  {Puig}}, \bibinfo {author} {\bibfnamefont {J.~A.}\ \bibnamefont {S\'anchez}},
  \bibinfo {author} {\bibfnamefont {E.}~\bibnamefont {Herrera}}, \bibinfo
  {author} {\bibfnamefont {I.}~\bibnamefont {Guillam\'on}}, \bibinfo {author}
  {\bibfnamefont {Z.}~\bibnamefont {Pribulov\'a}}, \bibinfo {author}
  {\bibfnamefont {J.}~\bibnamefont {Ka\ifmmode \check{c}\else
  \v{c}\fi{}mar\ifmmode~\check{c}\else \v{c}\fi{}\'{\i}k}}, \bibinfo {author}
  {\bibfnamefont {H.}~\bibnamefont {Suderow}}, \bibinfo {author} {\bibfnamefont
  {A.~B.}\ \bibnamefont {Kolton}},\ and\ \bibinfo {author} {\bibfnamefont
  {Y.}~\bibnamefont {Fasano}},\ }\bibfield  {title} {\bibinfo {title}
  {Anti-hyperuniform diluted vortex matter induced by correlated disorder},\
  }\href {https://doi.org/10.1103/PhysRevB.110.024108} {\bibfield  {journal}
  {\bibinfo  {journal} {Phys. Rev. B}\ }\textbf {\bibinfo {volume} {110}},\
  \bibinfo {pages} {024108} (\bibinfo {year} {2024})}\BibitemShut {NoStop}%
\bibitem [{\citenamefont {Leoni}\ \emph {et~al.}(2024)\citenamefont {Leoni},
  \citenamefont {O{\u{g}}uz},\ and\ \citenamefont
  {Franzese}}]{leoni2024emergence}%
  \BibitemOpen
  \bibfield  {author} {\bibinfo {author} {\bibfnamefont {F.}~\bibnamefont
  {Leoni}}, \bibinfo {author} {\bibfnamefont {E.~C.}\ \bibnamefont
  {O{\u{g}}uz}},\ and\ \bibinfo {author} {\bibfnamefont {G.}~\bibnamefont
  {Franzese}},\ }\bibfield  {title} {\bibinfo {title} {Emergence of disordered
  hyperuniformity in confined fluids and soft matter},\ }\href@noop {}
  {\bibfield  {journal} {\bibinfo  {journal} {arXiv preprint arXiv:2411.12393}\
  } (\bibinfo {year} {2024})}\BibitemShut {NoStop}%
\bibitem [{\citenamefont {Maher}\ and\ \citenamefont
  {Torquato}(2024{\natexlab{b}})}]{maher2024local}%
  \BibitemOpen
  \bibfield  {author} {\bibinfo {author} {\bibfnamefont {C.~E.}\ \bibnamefont
  {Maher}}\ and\ \bibinfo {author} {\bibfnamefont {S.}~\bibnamefont
  {Torquato}},\ }\bibfield  {title} {\bibinfo {title} {Local order metrics for
  many-particle systems across length scales},\ }\href@noop {} {\bibfield
  {journal} {\bibinfo  {journal} {{Physical Review Research}}\ }\textbf
  {\bibinfo {volume} {6}},\ \bibinfo {pages} {033262} (\bibinfo {year}
  {2024}{\natexlab{b}})}\BibitemShut {NoStop}%
\bibitem [{\citenamefont {Crampin}\ \emph {et~al.}(1999)\citenamefont
  {Crampin}, \citenamefont {Gaffney},\ and\ \citenamefont
  {Maini}}]{crampin1999reaction}%
  \BibitemOpen
  \bibfield  {author} {\bibinfo {author} {\bibfnamefont {E.~J.}\ \bibnamefont
  {Crampin}}, \bibinfo {author} {\bibfnamefont {E.~A.}\ \bibnamefont
  {Gaffney}},\ and\ \bibinfo {author} {\bibfnamefont {P.~K.}\ \bibnamefont
  {Maini}},\ }\bibfield  {title} {\bibinfo {title} {Reaction and diffusion on
  growing domains: scenarios for robust pattern formation},\ }\href@noop {}
  {\bibfield  {journal} {\bibinfo  {journal} {Bulletin of mathematical
  biology}\ }\textbf {\bibinfo {volume} {61}},\ \bibinfo {pages} {1093}
  (\bibinfo {year} {1999})}\BibitemShut {NoStop}%
\bibitem [{\citenamefont {Crampin}\ \emph {et~al.}(2002)\citenamefont
  {Crampin}, \citenamefont {Hackborn},\ and\ \citenamefont
  {Maini}}]{crampin2002pattern}%
  \BibitemOpen
  \bibfield  {author} {\bibinfo {author} {\bibfnamefont {E.~J.}\ \bibnamefont
  {Crampin}}, \bibinfo {author} {\bibfnamefont {W.~W.}\ \bibnamefont
  {Hackborn}},\ and\ \bibinfo {author} {\bibfnamefont {P.~K.}\ \bibnamefont
  {Maini}},\ }\bibfield  {title} {\bibinfo {title} {Pattern formation in
  reaction-diffusion models with nonuniform domain growth},\ }\href@noop {}
  {\bibfield  {journal} {\bibinfo  {journal} {Bulletin of mathematical
  biology}\ }\textbf {\bibinfo {volume} {64}},\ \bibinfo {pages} {747}
  (\bibinfo {year} {2002})}\BibitemShut {NoStop}%
\bibitem [{\citenamefont {Krause}\ \emph {et~al.}(2019)\citenamefont {Krause},
  \citenamefont {Ellis},\ and\ \citenamefont
  {Van~Gorder}}]{krause2019influence}%
  \BibitemOpen
  \bibfield  {author} {\bibinfo {author} {\bibfnamefont {A.~L.}\ \bibnamefont
  {Krause}}, \bibinfo {author} {\bibfnamefont {M.~A.}\ \bibnamefont {Ellis}},\
  and\ \bibinfo {author} {\bibfnamefont {R.~A.}\ \bibnamefont {Van~Gorder}},\
  }\bibfield  {title} {\bibinfo {title} {{Influence of curvature, growth, and
  anisotropy on the evolution of Turing patterns on growing manifolds}},\
  }\href@noop {} {\bibfield  {journal} {\bibinfo  {journal} {Bulletin of
  mathematical biology}\ }\textbf {\bibinfo {volume} {81}},\ \bibinfo {pages}
  {759} (\bibinfo {year} {2019})}\BibitemShut {NoStop}%
\bibitem [{\citenamefont {Ross}\ \emph {et~al.}(2017)\citenamefont {Ross},
  \citenamefont {Yates},\ and\ \citenamefont {Baker}}]{ross2017variable}%
  \BibitemOpen
  \bibfield  {author} {\bibinfo {author} {\bibfnamefont {R.~J.}\ \bibnamefont
  {Ross}}, \bibinfo {author} {\bibfnamefont {C.}~\bibnamefont {Yates}},\ and\
  \bibinfo {author} {\bibfnamefont {R.~E.}\ \bibnamefont {Baker}},\ }\bibfield
  {title} {\bibinfo {title} {Variable species densities are induced by volume
  exclusion interactions upon domain growth},\ }\href@noop {} {\bibfield
  {journal} {\bibinfo  {journal} {Physical Review E}\ }\textbf {\bibinfo
  {volume} {95}},\ \bibinfo {pages} {032416} (\bibinfo {year}
  {2017})}\BibitemShut {NoStop}%
\bibitem [{\citenamefont {Ross}\ \emph {et~al.}(2016)\citenamefont {Ross},
  \citenamefont {Baker},\ and\ \citenamefont {Yates}}]{ross2016domain}%
  \BibitemOpen
  \bibfield  {author} {\bibinfo {author} {\bibfnamefont {R.~J.}\ \bibnamefont
  {Ross}}, \bibinfo {author} {\bibfnamefont {R.~E.}\ \bibnamefont {Baker}},\
  and\ \bibinfo {author} {\bibfnamefont {C.~A.}\ \bibnamefont {Yates}},\
  }\bibfield  {title} {\bibinfo {title} {How domain growth is implemented
  determines the long-term behavior of a cell population through its effect on
  spatial correlations},\ }\href@noop {} {\bibfield  {journal} {\bibinfo
  {journal} {Physical Review E}\ }\textbf {\bibinfo {volume} {94}},\ \bibinfo
  {pages} {012408} (\bibinfo {year} {2016})}\BibitemShut {NoStop}%
\bibitem [{\citenamefont {Packard}\ and\ \citenamefont
  {Hochberg}(1977)}]{packard1977skin}%
  \BibitemOpen
  \bibfield  {author} {\bibinfo {author} {\bibfnamefont {A.}~\bibnamefont
  {Packard}}\ and\ \bibinfo {author} {\bibfnamefont {F.}~\bibnamefont
  {Hochberg}},\ }\bibfield  {title} {\bibinfo {title} {Skin patterning in
  octopus and other genera},\ }in\ \href@noop {} {\emph {\bibinfo {booktitle}
  {Symp. Zool. Soc. Lond}}},\ Vol.~\bibinfo {volume} {38}\ (\bibinfo {year}
  {1977})\ pp.\ \bibinfo {pages} {191--231}\BibitemShut {NoStop}%
\bibitem [{\citenamefont {Reiter}\ \emph {et~al.}(2018)\citenamefont {Reiter},
  \citenamefont {H{\"u}lsdunk}, \citenamefont {Woo}, \citenamefont
  {Lauterbach}, \citenamefont {Eberle}, \citenamefont {Akay}, \citenamefont
  {Longo}, \citenamefont {Meier-Credo}, \citenamefont {Kretschmer},
  \citenamefont {Langer} \emph {et~al.}}]{reiter2018elucidating}%
  \BibitemOpen
  \bibfield  {author} {\bibinfo {author} {\bibfnamefont {S.}~\bibnamefont
  {Reiter}}, \bibinfo {author} {\bibfnamefont {P.}~\bibnamefont
  {H{\"u}lsdunk}}, \bibinfo {author} {\bibfnamefont {T.}~\bibnamefont {Woo}},
  \bibinfo {author} {\bibfnamefont {M.~A.}\ \bibnamefont {Lauterbach}},
  \bibinfo {author} {\bibfnamefont {J.~S.}\ \bibnamefont {Eberle}}, \bibinfo
  {author} {\bibfnamefont {L.~A.}\ \bibnamefont {Akay}}, \bibinfo {author}
  {\bibfnamefont {A.}~\bibnamefont {Longo}}, \bibinfo {author} {\bibfnamefont
  {J.}~\bibnamefont {Meier-Credo}}, \bibinfo {author} {\bibfnamefont
  {F.}~\bibnamefont {Kretschmer}}, \bibinfo {author} {\bibfnamefont {J.~D.}\
  \bibnamefont {Langer}}, \emph {et~al.},\ }\bibfield  {title} {\bibinfo
  {title} {Elucidating the control and development of skin patterning in
  cuttlefish},\ }\href@noop {} {\bibfield  {journal} {\bibinfo  {journal}
  {Nature}\ }\textbf {\bibinfo {volume} {562}},\ \bibinfo {pages} {361}
  (\bibinfo {year} {2018})}\BibitemShut {NoStop}%
\bibitem [{\citenamefont {Messenger}(2001)}]{messenger2001cephalopod}%
  \BibitemOpen
  \bibfield  {author} {\bibinfo {author} {\bibfnamefont {J.~B.}\ \bibnamefont
  {Messenger}},\ }\bibfield  {title} {\bibinfo {title} {Cephalopod
  chromatophores: neurobiology and natural history},\ }\href@noop {} {\bibfield
   {journal} {\bibinfo  {journal} {Biological Reviews}\ }\textbf {\bibinfo
  {volume} {76}},\ \bibinfo {pages} {473} (\bibinfo {year} {2001})}\BibitemShut
  {NoStop}%
\bibitem [{\citenamefont {Packard}(2011)}]{packard2011squids}%
  \BibitemOpen
  \bibfield  {author} {\bibinfo {author} {\bibfnamefont {A.}~\bibnamefont
  {Packard}},\ }\bibfield  {title} {\bibinfo {title} {Squids old and young:
  Scale-free design for a simple billboard},\ }\href@noop {} {\bibfield
  {journal} {\bibinfo  {journal} {Optics \& Laser Technology}\ }\textbf
  {\bibinfo {volume} {43}},\ \bibinfo {pages} {302} (\bibinfo {year}
  {2011})}\BibitemShut {NoStop}%
\bibitem [{\citenamefont {Packard}(1982)}]{packard1982chromatophores}%
  \BibitemOpen
  \bibfield  {author} {\bibinfo {author} {\bibfnamefont {A.}~\bibnamefont
  {Packard}},\ }\bibfield  {title} {\bibinfo {title} {Morphogenesis of
  chromatophore patterns in cephalopods: Are morphological and physiological
  'units' the same?},\ }\href@noop {} {\bibfield  {journal} {\bibinfo
  {journal} {Malacologia}\ }\textbf {\bibinfo {volume} {23}},\ \bibinfo {pages}
  {193} (\bibinfo {year} {1982})}\BibitemShut {NoStop}%
\bibitem [{\citenamefont {Samuel}\ and\ \citenamefont
  {Patterson}(2002)}]{samuel2002intercapsular}%
  \BibitemOpen
  \bibfield  {author} {\bibinfo {author} {\bibfnamefont {V.~D.}\ \bibnamefont
  {Samuel}}\ and\ \bibinfo {author} {\bibfnamefont {J.}~\bibnamefont
  {Patterson}},\ }\bibfield  {title} {\bibinfo {title} {Intercapsular embryonic
  development of the big fin squid sepioteuthis lessoniana (loliginidae)},\
  }\href@noop {} {\bibfield  {journal} {\bibinfo  {journal} {Indian journal of
  marine sciences}\ }\textbf {\bibinfo {volume} {31}},\ \bibinfo {pages} {150}
  (\bibinfo {year} {2002})}\BibitemShut {NoStop}%
\bibitem [{\citenamefont {Chan}\ \emph {et~al.}(2022)\citenamefont {Chan},
  \citenamefont {Yan}, \citenamefont {Roan}, \citenamefont {Hsu}, \citenamefont
  {Tseng}, \citenamefont {Hsiao}, \citenamefont {Hsu},\ and\ \citenamefont
  {Chen}}]{chan2022skin}%
  \BibitemOpen
  \bibfield  {author} {\bibinfo {author} {\bibfnamefont {K.~Y.}\ \bibnamefont
  {Chan}}, \bibinfo {author} {\bibfnamefont {C.-C.~S.}\ \bibnamefont {Yan}},
  \bibinfo {author} {\bibfnamefont {H.-Y.}\ \bibnamefont {Roan}}, \bibinfo
  {author} {\bibfnamefont {S.-C.}\ \bibnamefont {Hsu}}, \bibinfo {author}
  {\bibfnamefont {T.-L.}\ \bibnamefont {Tseng}}, \bibinfo {author}
  {\bibfnamefont {C.-D.}\ \bibnamefont {Hsiao}}, \bibinfo {author}
  {\bibfnamefont {C.-P.}\ \bibnamefont {Hsu}},\ and\ \bibinfo {author}
  {\bibfnamefont {C.-H.}\ \bibnamefont {Chen}},\ }\bibfield  {title} {\bibinfo
  {title} {Skin cells undergo asynthetic fission to expand body surfaces in
  zebrafish},\ }\href@noop {} {\bibfield  {journal} {\bibinfo  {journal}
  {Nature}\ }\textbf {\bibinfo {volume} {605}},\ \bibinfo {pages} {119}
  (\bibinfo {year} {2022})}\BibitemShut {NoStop}%
\bibitem [{\citenamefont {Dekoninck}\ \emph {et~al.}(2020)\citenamefont
  {Dekoninck}, \citenamefont {Hannezo}, \citenamefont {Sifrim}, \citenamefont
  {Miroshnikova}, \citenamefont {Aragona}, \citenamefont {Malfait},
  \citenamefont {Gargouri}, \citenamefont {De~Neunheuser}, \citenamefont
  {Dubois}, \citenamefont {Voet} \emph {et~al.}}]{dekoninck2020defining}%
  \BibitemOpen
  \bibfield  {author} {\bibinfo {author} {\bibfnamefont {S.}~\bibnamefont
  {Dekoninck}}, \bibinfo {author} {\bibfnamefont {E.}~\bibnamefont {Hannezo}},
  \bibinfo {author} {\bibfnamefont {A.}~\bibnamefont {Sifrim}}, \bibinfo
  {author} {\bibfnamefont {Y.~A.}\ \bibnamefont {Miroshnikova}}, \bibinfo
  {author} {\bibfnamefont {M.}~\bibnamefont {Aragona}}, \bibinfo {author}
  {\bibfnamefont {M.}~\bibnamefont {Malfait}}, \bibinfo {author} {\bibfnamefont
  {S.}~\bibnamefont {Gargouri}}, \bibinfo {author} {\bibfnamefont
  {C.}~\bibnamefont {De~Neunheuser}}, \bibinfo {author} {\bibfnamefont
  {C.}~\bibnamefont {Dubois}}, \bibinfo {author} {\bibfnamefont
  {T.}~\bibnamefont {Voet}}, \emph {et~al.},\ }\bibfield  {title} {\bibinfo
  {title} {Defining the design principles of skin epidermis postnatal growth},\
  }\href@noop {} {\bibfield  {journal} {\bibinfo  {journal} {Cell}\ }\textbf
  {\bibinfo {volume} {181}},\ \bibinfo {pages} {604} (\bibinfo {year}
  {2020})}\BibitemShut {NoStop}%
\bibitem [{\citenamefont {Itzkovitz}\ \emph {et~al.}(2012)\citenamefont
  {Itzkovitz}, \citenamefont {Blat}, \citenamefont {Jacks}, \citenamefont
  {Clevers},\ and\ \citenamefont {van Oudenaarden}}]{itzkovitz2012optimality}%
  \BibitemOpen
  \bibfield  {author} {\bibinfo {author} {\bibfnamefont {S.}~\bibnamefont
  {Itzkovitz}}, \bibinfo {author} {\bibfnamefont {I.~C.}\ \bibnamefont {Blat}},
  \bibinfo {author} {\bibfnamefont {T.}~\bibnamefont {Jacks}}, \bibinfo
  {author} {\bibfnamefont {H.}~\bibnamefont {Clevers}},\ and\ \bibinfo {author}
  {\bibfnamefont {A.}~\bibnamefont {van Oudenaarden}},\ }\bibfield  {title}
  {\bibinfo {title} {Optimality in the development of intestinal crypts},\
  }\href@noop {} {\bibfield  {journal} {\bibinfo  {journal} {Cell}\ }\textbf
  {\bibinfo {volume} {148}},\ \bibinfo {pages} {608} (\bibinfo {year}
  {2012})}\BibitemShut {NoStop}%
\bibitem [{\citenamefont {Deryckere}\ \emph {et~al.}(2021)\citenamefont
  {Deryckere}, \citenamefont {Styfhals}, \citenamefont {Elagoz}, \citenamefont
  {Maes},\ and\ \citenamefont {Seuntjens}}]{deryckere2021identification}%
  \BibitemOpen
  \bibfield  {author} {\bibinfo {author} {\bibfnamefont {A.}~\bibnamefont
  {Deryckere}}, \bibinfo {author} {\bibfnamefont {R.}~\bibnamefont {Styfhals}},
  \bibinfo {author} {\bibfnamefont {A.~M.}\ \bibnamefont {Elagoz}}, \bibinfo
  {author} {\bibfnamefont {G.~E.}\ \bibnamefont {Maes}},\ and\ \bibinfo
  {author} {\bibfnamefont {E.}~\bibnamefont {Seuntjens}},\ }\bibfield  {title}
  {\bibinfo {title} {Identification of neural progenitor cells and their
  progeny reveals long distance migration in the developing octopus brain},\
  }\href@noop {} {\bibfield  {journal} {\bibinfo  {journal} {Elife}\ }\textbf
  {\bibinfo {volume} {10}},\ \bibinfo {pages} {e69161} (\bibinfo {year}
  {2021})}\BibitemShut {NoStop}%
\bibitem [{\citenamefont {Garc{\'\i}a-Moreno}\ and\ \citenamefont
  {Moln{\'a}r}(2020)}]{garcia2020impact}%
  \BibitemOpen
  \bibfield  {author} {\bibinfo {author} {\bibfnamefont {F.}~\bibnamefont
  {Garc{\'\i}a-Moreno}}\ and\ \bibinfo {author} {\bibfnamefont
  {Z.}~\bibnamefont {Moln{\'a}r}},\ }\bibfield  {title} {\bibinfo {title} {The
  impact of different modes of neuronal migration on brain evolution},\ }in\
  \href@noop {} {\emph {\bibinfo {booktitle} {Cellular migration and formation
  of axons and dendrites}}}\ (\bibinfo  {publisher} {Elsevier},\ \bibinfo
  {year} {2020})\ pp.\ \bibinfo {pages} {555--576}\BibitemShut {NoStop}%
\bibitem [{\citenamefont {Lenne}\ and\ \citenamefont
  {Trivedi}(2022)}]{lenne2022sculpting}%
  \BibitemOpen
  \bibfield  {author} {\bibinfo {author} {\bibfnamefont {P.-F.}\ \bibnamefont
  {Lenne}}\ and\ \bibinfo {author} {\bibfnamefont {V.}~\bibnamefont
  {Trivedi}},\ }\bibfield  {title} {\bibinfo {title} {Sculpting tissues by
  phase transitions},\ }\href@noop {} {\bibfield  {journal} {\bibinfo
  {journal} {Nature Communications}\ }\textbf {\bibinfo {volume} {13}},\
  \bibinfo {pages} {664} (\bibinfo {year} {2022})}\BibitemShut {NoStop}%
\bibitem [{\citenamefont {Coen}\ and\ \citenamefont
  {Cosgrove}(2023)}]{coen2023mechanics}%
  \BibitemOpen
  \bibfield  {author} {\bibinfo {author} {\bibfnamefont {E.}~\bibnamefont
  {Coen}}\ and\ \bibinfo {author} {\bibfnamefont {D.~J.}\ \bibnamefont
  {Cosgrove}},\ }\bibfield  {title} {\bibinfo {title} {The mechanics of plant
  morphogenesis},\ }\href@noop {} {\bibfield  {journal} {\bibinfo  {journal}
  {Science}\ }\textbf {\bibinfo {volume} {379}},\ \bibinfo {pages} {eade8055}
  (\bibinfo {year} {2023})}\BibitemShut {NoStop}%
\bibitem [{\citenamefont {Alim}\ \emph {et~al.}(2016)\citenamefont {Alim},
  \citenamefont {Armon}, \citenamefont {Shraiman},\ and\ \citenamefont
  {Boudaoud}}]{alim2016leaf}%
  \BibitemOpen
  \bibfield  {author} {\bibinfo {author} {\bibfnamefont {K.}~\bibnamefont
  {Alim}}, \bibinfo {author} {\bibfnamefont {S.}~\bibnamefont {Armon}},
  \bibinfo {author} {\bibfnamefont {B.~I.}\ \bibnamefont {Shraiman}},\ and\
  \bibinfo {author} {\bibfnamefont {A.}~\bibnamefont {Boudaoud}},\ }\bibfield
  {title} {\bibinfo {title} {Leaf growth is conformal},\ }\href@noop {}
  {\bibfield  {journal} {\bibinfo  {journal} {Physical biology}\ }\textbf
  {\bibinfo {volume} {13}},\ \bibinfo {pages} {05LT01} (\bibinfo {year}
  {2016})}\BibitemShut {NoStop}%
\bibitem [{\citenamefont {Hidalgo}\ \emph {et~al.}(2014)\citenamefont
  {Hidalgo}, \citenamefont {Grilli}, \citenamefont {Suweis}, \citenamefont
  {Munoz}, \citenamefont {Banavar},\ and\ \citenamefont
  {Maritan}}]{hidalgo2014information}%
  \BibitemOpen
  \bibfield  {author} {\bibinfo {author} {\bibfnamefont {J.}~\bibnamefont
  {Hidalgo}}, \bibinfo {author} {\bibfnamefont {J.}~\bibnamefont {Grilli}},
  \bibinfo {author} {\bibfnamefont {S.}~\bibnamefont {Suweis}}, \bibinfo
  {author} {\bibfnamefont {M.~A.}\ \bibnamefont {Munoz}}, \bibinfo {author}
  {\bibfnamefont {J.~R.}\ \bibnamefont {Banavar}},\ and\ \bibinfo {author}
  {\bibfnamefont {A.}~\bibnamefont {Maritan}},\ }\bibfield  {title} {\bibinfo
  {title} {Information-based fitness and the emergence of criticality in living
  systems},\ }\href@noop {} {\bibfield  {journal} {\bibinfo  {journal}
  {Proceedings of the National Academy of Sciences}\ }\textbf {\bibinfo
  {volume} {111}},\ \bibinfo {pages} {10095} (\bibinfo {year}
  {2014})}\BibitemShut {NoStop}%
\bibitem [{\citenamefont {Fruleux}\ \emph {et~al.}(2024)\citenamefont
  {Fruleux}, \citenamefont {Hong}, \citenamefont {Roeder}, \citenamefont {Li},\
  and\ \citenamefont {Boudaoud}}]{fruleux2024growth}%
  \BibitemOpen
  \bibfield  {author} {\bibinfo {author} {\bibfnamefont {A.}~\bibnamefont
  {Fruleux}}, \bibinfo {author} {\bibfnamefont {L.}~\bibnamefont {Hong}},
  \bibinfo {author} {\bibfnamefont {A.~H.}\ \bibnamefont {Roeder}}, \bibinfo
  {author} {\bibfnamefont {C.-B.}\ \bibnamefont {Li}},\ and\ \bibinfo {author}
  {\bibfnamefont {A.}~\bibnamefont {Boudaoud}},\ }\bibfield  {title} {\bibinfo
  {title} {Growth couples temporal and spatial fluctuations of tissue
  properties during morphogenesis},\ }\href@noop {} {\bibfield  {journal}
  {\bibinfo  {journal} {Proceedings of the National Academy of Sciences}\
  }\textbf {\bibinfo {volume} {121}},\ \bibinfo {pages} {e2318481121} (\bibinfo
  {year} {2024})}\BibitemShut {NoStop}%
\bibitem [{\citenamefont {Gould}\ and\ \citenamefont
  {Lewontin}(1979)}]{gould1979spandrels}%
  \BibitemOpen
  \bibfield  {author} {\bibinfo {author} {\bibfnamefont {S.~J.}\ \bibnamefont
  {Gould}}\ and\ \bibinfo {author} {\bibfnamefont {R.~C.}\ \bibnamefont
  {Lewontin}},\ }\bibfield  {title} {\bibinfo {title} {The spandrels of san
  marco and the panglossian paradigm: a critique of the adaptationist
  programme},\ }\href@noop {} {\bibfield  {journal} {\bibinfo  {journal} {Proc.
  R. Soc. Lond. B}\ }\textbf {\bibinfo {volume} {205}},\ \bibinfo {pages} {581}
  (\bibinfo {year} {1979})}\BibitemShut {NoStop}%
\bibitem [{\citenamefont {Wartlick}\ \emph {et~al.}(2011)\citenamefont
  {Wartlick}, \citenamefont {Mumcu}, \citenamefont {Kicheva}, \citenamefont
  {Bittig}, \citenamefont {Seum}, \citenamefont {J{\"u}licher},\ and\
  \citenamefont {Gonzalez-Gaitan}}]{wartlick2011dynamics}%
  \BibitemOpen
  \bibfield  {author} {\bibinfo {author} {\bibfnamefont {O.}~\bibnamefont
  {Wartlick}}, \bibinfo {author} {\bibfnamefont {P.}~\bibnamefont {Mumcu}},
  \bibinfo {author} {\bibfnamefont {A.}~\bibnamefont {Kicheva}}, \bibinfo
  {author} {\bibfnamefont {T.}~\bibnamefont {Bittig}}, \bibinfo {author}
  {\bibfnamefont {C.}~\bibnamefont {Seum}}, \bibinfo {author} {\bibfnamefont
  {F.}~\bibnamefont {J{\"u}licher}},\ and\ \bibinfo {author} {\bibfnamefont
  {M.}~\bibnamefont {Gonzalez-Gaitan}},\ }\bibfield  {title} {\bibinfo {title}
  {Dynamics of dpp signaling and proliferation control},\ }\href@noop {}
  {\bibfield  {journal} {\bibinfo  {journal} {Science}\ }\textbf {\bibinfo
  {volume} {331}},\ \bibinfo {pages} {1154} (\bibinfo {year}
  {2011})}\BibitemShut {NoStop}%
\bibitem [{\citenamefont {Averbukh}\ \emph {et~al.}(2014)\citenamefont
  {Averbukh}, \citenamefont {Ben-Zvi}, \citenamefont {Mishra},\ and\
  \citenamefont {Barkai}}]{averbukh2014scaling}%
  \BibitemOpen
  \bibfield  {author} {\bibinfo {author} {\bibfnamefont {I.}~\bibnamefont
  {Averbukh}}, \bibinfo {author} {\bibfnamefont {D.}~\bibnamefont {Ben-Zvi}},
  \bibinfo {author} {\bibfnamefont {S.}~\bibnamefont {Mishra}},\ and\ \bibinfo
  {author} {\bibfnamefont {N.}~\bibnamefont {Barkai}},\ }\bibfield  {title}
  {\bibinfo {title} {Scaling morphogen gradients during tissue growth by a cell
  division rule},\ }\href@noop {} {\bibfield  {journal} {\bibinfo  {journal}
  {Development}\ }\textbf {\bibinfo {volume} {141}},\ \bibinfo {pages} {2150}
  (\bibinfo {year} {2014})}\BibitemShut {NoStop}%
\bibitem [{\citenamefont {Aguilar-Hidalgo}\ \emph {et~al.}(2018)\citenamefont
  {Aguilar-Hidalgo}, \citenamefont {Werner}, \citenamefont {Wartlick},
  \citenamefont {Gonz{\'a}lez-Gait{\'a}n}, \citenamefont {Friedrich},\ and\
  \citenamefont {J{\"u}licher}}]{aguilar2018critical}%
  \BibitemOpen
  \bibfield  {author} {\bibinfo {author} {\bibfnamefont {D.}~\bibnamefont
  {Aguilar-Hidalgo}}, \bibinfo {author} {\bibfnamefont {S.}~\bibnamefont
  {Werner}}, \bibinfo {author} {\bibfnamefont {O.}~\bibnamefont {Wartlick}},
  \bibinfo {author} {\bibfnamefont {M.}~\bibnamefont
  {Gonz{\'a}lez-Gait{\'a}n}}, \bibinfo {author} {\bibfnamefont {B.~M.}\
  \bibnamefont {Friedrich}},\ and\ \bibinfo {author} {\bibfnamefont
  {F.}~\bibnamefont {J{\"u}licher}},\ }\bibfield  {title} {\bibinfo {title}
  {Critical point in self-organized tissue growth},\ }\href@noop {} {\bibfield
  {journal} {\bibinfo  {journal} {Physical review letters}\ }\textbf {\bibinfo
  {volume} {120}},\ \bibinfo {pages} {198102} (\bibinfo {year}
  {2018})}\BibitemShut {NoStop}%
\bibitem [{\citenamefont {Boudaoud}(2010)}]{boudaoud2010introduction}%
  \BibitemOpen
  \bibfield  {author} {\bibinfo {author} {\bibfnamefont {A.}~\bibnamefont
  {Boudaoud}},\ }\bibfield  {title} {\bibinfo {title} {An introduction to the
  mechanics of morphogenesis for plant biologists},\ }\href@noop {} {\bibfield
  {journal} {\bibinfo  {journal} {Trends in plant science}\ }\textbf {\bibinfo
  {volume} {15}},\ \bibinfo {pages} {353} (\bibinfo {year} {2010})}\BibitemShut
  {NoStop}%
\bibitem [{\citenamefont {Uyttewaal}\ \emph {et~al.}(2012)\citenamefont
  {Uyttewaal}, \citenamefont {Burian}, \citenamefont {Alim}, \citenamefont
  {Landrein}, \citenamefont {Borowska-Wykret}, \citenamefont {Dedieu},
  \citenamefont {Peaucelle}, \citenamefont {Ludynia}, \citenamefont {Traas},
  \citenamefont {Boudaoud} \emph {et~al.}}]{uyttewaal2012mechanical}%
  \BibitemOpen
  \bibfield  {author} {\bibinfo {author} {\bibfnamefont {M.}~\bibnamefont
  {Uyttewaal}}, \bibinfo {author} {\bibfnamefont {A.}~\bibnamefont {Burian}},
  \bibinfo {author} {\bibfnamefont {K.}~\bibnamefont {Alim}}, \bibinfo {author}
  {\bibfnamefont {B.}~\bibnamefont {Landrein}}, \bibinfo {author}
  {\bibfnamefont {D.}~\bibnamefont {Borowska-Wykret}}, \bibinfo {author}
  {\bibfnamefont {A.}~\bibnamefont {Dedieu}}, \bibinfo {author} {\bibfnamefont
  {A.}~\bibnamefont {Peaucelle}}, \bibinfo {author} {\bibfnamefont
  {M.}~\bibnamefont {Ludynia}}, \bibinfo {author} {\bibfnamefont
  {J.}~\bibnamefont {Traas}}, \bibinfo {author} {\bibfnamefont
  {A.}~\bibnamefont {Boudaoud}}, \emph {et~al.},\ }\bibfield  {title} {\bibinfo
  {title} {Mechanical stress acts via katanin to amplify differences in growth
  rate between adjacent cells in arabidopsis},\ }\href@noop {} {\bibfield
  {journal} {\bibinfo  {journal} {Cell}\ }\textbf {\bibinfo {volume} {149}},\
  \bibinfo {pages} {439} (\bibinfo {year} {2012})}\BibitemShut {NoStop}%
\bibitem [{\citenamefont {Hamant}\ \emph {et~al.}(2008)\citenamefont {Hamant},
  \citenamefont {Heisler}, \citenamefont {Jonsson}, \citenamefont {Krupinski},
  \citenamefont {Uyttewaal}, \citenamefont {Bokov}, \citenamefont {Corson},
  \citenamefont {Sahlin}, \citenamefont {Boudaoud}, \citenamefont {Meyerowitz}
  \emph {et~al.}}]{hamant2008developmental}%
  \BibitemOpen
  \bibfield  {author} {\bibinfo {author} {\bibfnamefont {O.}~\bibnamefont
  {Hamant}}, \bibinfo {author} {\bibfnamefont {M.~G.}\ \bibnamefont {Heisler}},
  \bibinfo {author} {\bibfnamefont {H.}~\bibnamefont {Jonsson}}, \bibinfo
  {author} {\bibfnamefont {P.}~\bibnamefont {Krupinski}}, \bibinfo {author}
  {\bibfnamefont {M.}~\bibnamefont {Uyttewaal}}, \bibinfo {author}
  {\bibfnamefont {P.}~\bibnamefont {Bokov}}, \bibinfo {author} {\bibfnamefont
  {F.}~\bibnamefont {Corson}}, \bibinfo {author} {\bibfnamefont
  {P.}~\bibnamefont {Sahlin}}, \bibinfo {author} {\bibfnamefont
  {A.}~\bibnamefont {Boudaoud}}, \bibinfo {author} {\bibfnamefont {E.~M.}\
  \bibnamefont {Meyerowitz}}, \emph {et~al.},\ }\bibfield  {title} {\bibinfo
  {title} {Developmental patterning by mechanical signals in arabidopsis},\
  }\href@noop {} {\bibfield  {journal} {\bibinfo  {journal} {Science}\ }\textbf
  {\bibinfo {volume} {322}},\ \bibinfo {pages} {1650} (\bibinfo {year}
  {2008})}\BibitemShut {NoStop}%
\bibitem [{\citenamefont {Shraiman}(2005)}]{shraiman2005mechanical}%
  \BibitemOpen
  \bibfield  {author} {\bibinfo {author} {\bibfnamefont {B.~I.}\ \bibnamefont
  {Shraiman}},\ }\bibfield  {title} {\bibinfo {title} {Mechanical feedback as a
  possible regulator of tissue growth},\ }\href@noop {} {\bibfield  {journal}
  {\bibinfo  {journal} {Proceedings of the National Academy of Sciences}\
  }\textbf {\bibinfo {volume} {102}},\ \bibinfo {pages} {3318} (\bibinfo {year}
  {2005})}\BibitemShut {NoStop}%
\bibitem [{\citenamefont {Lecuit}\ and\ \citenamefont
  {Le~Goff}(2007)}]{lecuit2007orchestrating}%
  \BibitemOpen
  \bibfield  {author} {\bibinfo {author} {\bibfnamefont {T.}~\bibnamefont
  {Lecuit}}\ and\ \bibinfo {author} {\bibfnamefont {L.}~\bibnamefont
  {Le~Goff}},\ }\bibfield  {title} {\bibinfo {title} {Orchestrating size and
  shape during morphogenesis},\ }\href@noop {} {\bibfield  {journal} {\bibinfo
  {journal} {Nature}\ }\textbf {\bibinfo {volume} {450}},\ \bibinfo {pages}
  {189} (\bibinfo {year} {2007})}\BibitemShut {NoStop}%
\bibitem [{\citenamefont {Salis}\ \emph {et~al.}(2021)\citenamefont {Salis},
  \citenamefont {Roux}, \citenamefont {Huang}, \citenamefont {Marcionetti},
  \citenamefont {Mouginot}, \citenamefont {Reynaud}, \citenamefont {Salles},
  \citenamefont {Salamin}, \citenamefont {Pujol}, \citenamefont {Parichy} \emph
  {et~al.}}]{salis2021thyroid}%
  \BibitemOpen
  \bibfield  {author} {\bibinfo {author} {\bibfnamefont {P.}~\bibnamefont
  {Salis}}, \bibinfo {author} {\bibfnamefont {N.}~\bibnamefont {Roux}},
  \bibinfo {author} {\bibfnamefont {D.}~\bibnamefont {Huang}}, \bibinfo
  {author} {\bibfnamefont {A.}~\bibnamefont {Marcionetti}}, \bibinfo {author}
  {\bibfnamefont {P.}~\bibnamefont {Mouginot}}, \bibinfo {author}
  {\bibfnamefont {M.}~\bibnamefont {Reynaud}}, \bibinfo {author} {\bibfnamefont
  {O.}~\bibnamefont {Salles}}, \bibinfo {author} {\bibfnamefont
  {N.}~\bibnamefont {Salamin}}, \bibinfo {author} {\bibfnamefont
  {B.}~\bibnamefont {Pujol}}, \bibinfo {author} {\bibfnamefont {D.~M.}\
  \bibnamefont {Parichy}}, \emph {et~al.},\ }\bibfield  {title} {\bibinfo
  {title} {Thyroid hormones regulate the formation and environmental plasticity
  of white bars in clownfishes},\ }\href@noop {} {\bibfield  {journal}
  {\bibinfo  {journal} {Proceedings of the National Academy of Sciences}\
  }\textbf {\bibinfo {volume} {118}},\ \bibinfo {pages} {e2101634118} (\bibinfo
  {year} {2021})}\BibitemShut {NoStop}%
\bibitem [{\citenamefont {Butler-Struben}\ \emph {et~al.}(2018)\citenamefont
  {Butler-Struben}, \citenamefont {Brophy}, \citenamefont {Johnson},\ and\
  \citenamefont {Crook}}]{butler2018vivo}%
  \BibitemOpen
  \bibfield  {author} {\bibinfo {author} {\bibfnamefont {H.~M.}\ \bibnamefont
  {Butler-Struben}}, \bibinfo {author} {\bibfnamefont {S.~M.}\ \bibnamefont
  {Brophy}}, \bibinfo {author} {\bibfnamefont {N.~A.}\ \bibnamefont
  {Johnson}},\ and\ \bibinfo {author} {\bibfnamefont {R.~J.}\ \bibnamefont
  {Crook}},\ }\bibfield  {title} {\bibinfo {title} {In vivo recording of neural
  and behavioral correlates of anesthesia induction, reversal, and euthanasia
  in cephalopod molluscs},\ }\href@noop {} {\bibfield  {journal} {\bibinfo
  {journal} {Frontiers in physiology}\ }\textbf {\bibinfo {volume} {9}},\
  \bibinfo {pages} {326068} (\bibinfo {year} {2018})}\BibitemShut {NoStop}%
\bibitem [{\citenamefont {Woo}\ \emph {et~al.}(2023)\citenamefont {Woo},
  \citenamefont {Liang}, \citenamefont {Evans}, \citenamefont {Fernandez},
  \citenamefont {Kretschmer}, \citenamefont {Reiter},\ and\ \citenamefont
  {Laurent}}]{woo2023dynamics}%
  \BibitemOpen
  \bibfield  {author} {\bibinfo {author} {\bibfnamefont {T.}~\bibnamefont
  {Woo}}, \bibinfo {author} {\bibfnamefont {X.}~\bibnamefont {Liang}}, \bibinfo
  {author} {\bibfnamefont {D.~A.}\ \bibnamefont {Evans}}, \bibinfo {author}
  {\bibfnamefont {O.}~\bibnamefont {Fernandez}}, \bibinfo {author}
  {\bibfnamefont {F.}~\bibnamefont {Kretschmer}}, \bibinfo {author}
  {\bibfnamefont {S.}~\bibnamefont {Reiter}},\ and\ \bibinfo {author}
  {\bibfnamefont {G.}~\bibnamefont {Laurent}},\ }\bibfield  {title} {\bibinfo
  {title} {The dynamics of pattern matching in camouflaging cuttlefish},\
  }\href@noop {} {\bibfield  {journal} {\bibinfo  {journal} {Nature}\ }\textbf
  {\bibinfo {volume} {619}},\ \bibinfo {pages} {122} (\bibinfo {year}
  {2023})}\BibitemShut {NoStop}%
\bibitem [{\citenamefont {Ronneberger}\ \emph {et~al.}(2015)\citenamefont
  {Ronneberger}, \citenamefont {Fischer},\ and\ \citenamefont
  {Brox}}]{ronneberger2015u}%
  \BibitemOpen
  \bibfield  {author} {\bibinfo {author} {\bibfnamefont {O.}~\bibnamefont
  {Ronneberger}}, \bibinfo {author} {\bibfnamefont {P.}~\bibnamefont
  {Fischer}},\ and\ \bibinfo {author} {\bibfnamefont {T.}~\bibnamefont
  {Brox}},\ }\href@noop {} {\emph {\bibinfo {title} {Medical image computing
  and computer-assisted intervention--MICCAI 2015: 18th international
  conference, Munich, Germany, October 5-9, 2015, proceedings, part III 18}}}\
  (\bibinfo {year} {2015})\ pp.\ \bibinfo {pages} {234--241}\BibitemShut
  {NoStop}%
\bibitem [{\citenamefont {Bhattacharjee}\ and\ \citenamefont
  {Seno}(2001)}]{bhattacharjee2001measure}%
  \BibitemOpen
  \bibfield  {author} {\bibinfo {author} {\bibfnamefont {S.~M.}\ \bibnamefont
  {Bhattacharjee}}\ and\ \bibinfo {author} {\bibfnamefont {F.}~\bibnamefont
  {Seno}},\ }\bibfield  {title} {\bibinfo {title} {A measure of data collapse
  for scaling},\ }\href@noop {} {\bibfield  {journal} {\bibinfo  {journal}
  {Journal of Physics A: Mathematical and General}\ }\textbf {\bibinfo {volume}
  {34}},\ \bibinfo {pages} {6375} (\bibinfo {year} {2001})}\BibitemShut
  {NoStop}%
\end{thebibliography}%


\end{document}